\documentclass[twocolumn,tighten]{aastex62}
\usepackage{verbatim}
\usepackage{color}
\usepackage{amsmath}
\usepackage{graphicx}
\usepackage{threeparttable}

\newcommand{\kepler}{\textit{Kepler} }

\newcommand{\bjdtdb}{\ensuremath{\rm {BJD_{TDB}}}}

\bibpunct[ ]{(}{)}{;}{}{}{,}
\shorttitle{KOI-964 \kepler phase curve}
\shortauthors{Wong et~al.}

\begin{document}
\title{The full \kepler phase curve of the eclipsing hot white dwarf binary system KOI-964}
\correspondingauthor{Ian Wong}
\email{iwong@mit.edu}

\author[0000-0001-9665-8429]{Ian~Wong}
\altaffiliation{51 Pegasi b Fellow}
\affil{Department of Earth, Atmospheric and Planetary Sciences, Massachusetts Institute of Technology,
Cambridge, MA 02139, USA}

\author[0000-0002-1836-3120]{Avi~Shporer}
\affil{Department of Physics and Kavli Institute for Astrophysics and Space Research, Massachusetts Institute of Technology, Cambridge, MA 02139, USA}

\author[0000-0002-7733-4522]{Juliette~C.~Becker}
\altaffiliation{51 Pegasi b Fellow}
\affil{Division of Geological and Planetary Sciences, California Institute of Technology, Pasadena, CA 91125, USA}

\author[0000-0003-3504-5316]{Benjamin~J.~Fulton}
\affil{Caltech/IPAC-NASA Exoplanet Science Institute, Pasadena, CA 91106, USA}

\author[0000-0002-2580-3614]{Travis~A.~Berger}
\affil{Institute for Astronomy, University of Hawaii, Honolulu, HI 96822, USA}

\author[0000-0001-9194-2084]{Nevin~N.~Weinberg}
\affil{Department of Physics and Kavli Institute for Astrophysics and Space Research, Massachusetts Institute of Technology, Cambridge, MA 02139, USA}

\author[0000-0001-5611-1349]{Phil~Arras}
\affil{Department of Astronomy, University of Virginia, Charlottesville, VA 22904, USA}

\author[0000-0001-8638-0320]{Andrew~W.~Howard}
\affil{Institute for Astronomy, University of Hawaii, Honolulu, HI 96822, USA}

\author[0000-0001-5578-1498]{Bj{\" o}rn~Benneke}
\affiliation{Department of Physics and Institute for Research on Exoplanets, Universit{\' e} de Montr{\' e}al, Montr{\' e}al, QC, Canada}

\begin{abstract}
We analyze the full \kepler phase curve of KOI-964, a binary system consisting of a hot white dwarf on an eclipsing orbit around an A-type host star. Using all 18 quarters of long-cadence photometry, we carry out a joint light curve fit and obtain improved phase curve amplitudes, occultation depths, orbital parameters, and transit ephemeris over the previous results of \citet{carter2011}. A periodogram of the residuals from the phase curve fit reveals an additional stellar variability signal from the host star with a characteristic period of $0.620276\pm0.000011$~days and a full amplitude of $24\pm2$~ppm. We also present new Keck/HIRES radial velocity observations which we use to measure the orbit and obtain a mass ratio of $q=0.106\pm0.012$. Combining this measurement with the results of a stellar isochrone analysis, we find that the masses of the host star and white dwarf companion are $2.23\pm0.12\,M_{\Sun}$ and $0.236^{+0.028}_{-0.027}\,M_{\Sun}$, respectively. The effective temperatures of the two components are $9940^{+260}_{-230}$~K and $15080\pm400$~K, respectively, and we determine the age of the system to be $0.21^{+0.11}_{-0.08}$~Gyr. We use the measured system properties to compute predicted phase curve amplitudes and find that while the measured Doppler boosting and mutual illumination components agree well with theory, the ellipsoidal distortion amplitude is significantly underestimated. We detail possible explanations for this discrepancy, including interactions between the dynamical tide of the host star and the tidal bulge and possible non-synchronous rotation of the host star.
\end{abstract}


\section{Introduction}\label{sec:intro}

Close-in binary systems often display periodic brightness modulations. At visible wavelengths, these phase curves are shaped by changes in the illuminated fraction of the orbiting companion's observer-facing hemisphere, as well as photometric variations induced by the mutual gravitational interaction. The amplitudes of various contributions to the overall phase curve signal can be calculated using theoretical models of the mutual illumination and stellar tidal distortion, which ultimately depend on fundamental system properties such as the brightness temperatures, orbital separation, and mass ratio \citep[see the review by][]{shporer2017}. Conversely, precise measurements of the phase curve from long-baseline photometry yield constraints on these fundamental properties. Phase curve observations are therefore an important tool for studying binary systems and serve as a powerful empirical test of our understanding of stellar astrophysics. 

There are two separate gravitational processes that can be manifested in visible-light phase curves. The first is known as ellipsoidal distortion, where the binary companions raise tidal bulges on each other's surfaces. This produces a modulation in the sky-projected areas of the two components, resulting in a brightness modulation that comes to maximum at quadrature \citep[e.g.,][]{morris1985,morris1993,pfahl2008,jackson2012}. Setting the zeropoint of orbital phase at inferior conjunction, this produces a signal at the first harmonic of the cosine of the orbital phase. The second process is called beaming or Doppler boosting. The radial velocities (RVs) induced by the mutual gravitational interaction lead to periodic blue- and red-shifting of the stellar spectra as well as modulations in photon emission rate in the observer's direction \citep[e.g.,][]{shakura1987,loeb2003,zucker2007,shporer2010}. The resultant photometric variation comes to maximum and minimum at the two quadratures, yielding a signal at the sine of the orbital phase. 

Detections of the ellipsoidal distortion and Doppler boosting signals provide independent measurements of the planet-star mass ratio. \citet{carter2011} previously carried out a detailed phase curve study of the binary system KOI-964, using the available \kepler data at the time. The KOI-964 system consists of a young low mass $T\sim15000$~K white dwarf on a 3.273-day orbit around a 2.3~$M_{\Sun}$ A-type star. They found that the mass ratios derived from the measured amplitudes of the Doppler boosting and ellipsoidal distortion phase curve components were discrepant, indicating that the physical description of one or both of these gravity-induced phase curve components is incomplete. 

In this work, we revisit the KOI-964 system and carry out an analysis of the full 4-year \kepler phase curve. We utilize new Keck/HIRES RV observations to measure the true mass ratio and compare it with the values derived from the measured phase curve amplitudes. Our observations and data analysis methodologies are described in Sections~\ref{sec:obs} and \ref{sec:data}. We present the results of our RV analysis and light curve fitting in Section~\ref{sec:res}. Section~\ref{sec:dis} provides a detailed discussion of the various phase components and a comparison between the predicted phase curve amplitudes calculated using the RV-derived mass ratio and those measured from our joint fit. We conclude with a brief summary in Section~\ref{sec:sum}.

\section{Observations}\label{sec:obs}

\subsection{\kepler photometry}\label{subsec:phot}
In our phase curve analysis of KOI-964, we utilize all available Single Aperture Photometry (SAP) data for the system taken during quarters zero through seventeen (Q0-Q17).  Most of the quarters contain $\sim$90~days of data, with the exception of Q0 ($\sim$10~d), Q1 ($\sim$33~d), and Q17 ($\sim$32~d). The continuous long-cadence photometric observations with 30-minute integrated exposures were interrupted by monthly data downlinks, quarterly field rotations, as well as other occasional cessations in data collection.

The official \kepler data processing pipeline \citep[e.g.,][]{jenkins2010,jenkins2017} provides quality flags that indicate when certain spacecraft actions occurred or when non-nominal operation may have yielded unreliable photometry. We discard all exposures with a non-zero quality flag value. We then apply a 16-point-wide ($\sim$8-hour) moving median filter to the photometric series, with transits and secondary eclipses masked, and remove 3$\sigma$ outliers. 

Inspection of the raw SAP light curves reveals clear periodic flux variations attributable to the astrophysical phase curve signal, as well as long-term low-frequency brightness modulations that indicate the presence of instrumental systematics (correlated noise). We also find occasional flux ramps lasting up to several days immediately following gaps in data collection, such as those corresponding to data downlinks. Detrending these ramp-like features in addition to the other long-term systematic trends requires additional systematics parameters and often incurs correlations between the systematics and astrophysical parameters. We therefore choose to consistently trim away five days worth of data after each gap. An exception is the short first quarter (Q0), since removing five days from the time series would leave just over one orbital period worth of data, with a single primary and secondary eclipse. Without a pair of primary or secondary eclipses to self-consistently constrain the orbital period, this segment would be insufficient to reliably compute the best-fit systematics model in a simultaneous fit with the astrophysical phase curve model (see below).

The data gaps and trimming divide each quarter's time series into discrete segments. In our initial analysis of the KOI-964 phase curve, we carry out individual fits of each segment separately in order to optimize the systematics model prior to the joint fit. Information on the full list of segments analyzed in this work is given in Table~\ref{tab:obs}.

\begin{deluxetable}{ccrrrr}
\tablewidth{0pc}
\tabletypesize{\scriptsize}
\tablecaption{
    \kepler  Observation Details
    \label{tab:obs}
}
\tablehead{
    \colhead{Quarter} &
    \colhead{Segment}                     &
    \colhead{$t_{\mathrm{start}}$\textsuperscript{a} }  &
    \colhead{$t_{\mathrm{end}}$\textsuperscript{a}} &
    \colhead{$n_{\mathrm{exp}}$\textsuperscript{b}} &
    \colhead{ Order\textsuperscript{c}}
}
\startdata
0 & 0 & -46.441 & -36.775 & 452 & 3 \\
1 & 0 & -30.481 & -2.017 & 1295 & 5 \\
2 & 0 & 7.771 & 14.515 & 304 & 3 \\
 & 1 & 21.728 & 33.293 & 524 & 2 \\
 & 2 & 38.361 & 56.404 & 787 & 2 \\
 & 3 & 69.399 & 79.166 & 434 & 2 \\
 & 4 & 84.193 & 91.467 & 286 & 1 \\
3 & 0\textsuperscript{d} & 98.231 & 123.547 & 1094 & 6 \\
 & 1\textsuperscript{d} & 129.452 & 154.441 & 1088 & 7 \\
 & 2 & 161.552 & 182.496 & 925 & 4 \\
4 & 0 & 190.383 & 204.727 & 650 & 6 \\
 & 1 & 211.266 & 216.415 & 223 & 1 \\
 & 2 & 221.503 & 229.840 & 388 & 1 \\
 & 3 & 238.831 & 275.203 & 1647 & 7 \\
5 & 0\textsuperscript{d} & 281.497 & 308.000 & 1190 & 6 \\
 & 1 & 314.294 & 324.961 & 480 & 5 \\
 & 2 & 340.041 & 371.162 & 1397 & 6 \\
6 & 0 & 377.476 & 399.525 & 987 & 4 \\
 & 1 & 405.389 & 431.279 & 1151 & 9 \\
 & 2 & 437.245 & 462.296 & 1112 & 8 \\
7 & 0\textsuperscript{d} & 468.181 & 493.273 & 1087 & 6 \\
 & 1\textsuperscript{d} & 499.076 & 523.227 & 1055 & 8 \\
 & 2 & 529.153 & 552.508 & 991 & 6 \\
8 & 0 & 573.370 & 594.048 & 947 & 6 \\
 & 1 & 601.793 & 635.345 & 1528 & 9 \\
9 & 0 & 646.523 & 677.910 & 1420 & 6 \\
 & 1 & 683.631 & 707.110 & 1066 & 6 \\
 & 2 & 712.791 & 719.636 & 312 & 2 \\
 & 3 & 725.583 & 738.926 & 614 & 3 \\
10 & 0 & 744.852 & 769.945 & 1135 & 6 \\
 & 1 & 775.809 & 802.230 & 1165 & 5 \\
 & 2 & 808.094 & 833.268 & 1104 & 7 \\
11 & 0\textsuperscript{d} & 839.276 & 865.266 & 1172 & 6 \\
 & 1\textsuperscript{d} & 871.069 & 896.222 & 1130 & 14 \\
 & 2 & 910.362 & 931.326 & 911 & 4 \\
12 & 0 & 937.415 & 949.675 & 578 & 4 \\
 & 1\textsuperscript{d} & 964.653 & 986.987 & 1018 & 6 \\
 & 2 & 1003.068 & 1015.022 & 566 & 2 \\
13 & 0 & 1020.764 & 1047.982 & 1237 & 4 \\
 & 1 & 1053.724 & 1077.919 & 1074 & 6 \\
 & 2 & 1083.865 & 1106.057 & 1019 & 6 \\
14 & 0 & 1112.146 & 1122.526 & 466 & 3 \\
 & 1 & 1144.166 & 1169.299 & 1150 & 4 \\
 & 2 & 1175.184 & 1204.322 & 1322 & 6 \\
 15 & 0\textsuperscript{d} & 1211.494 & 1237.301 & 1161 & 16 \\
 & 1 & 1256.406 & 1268.379 & 539 & 3 \\
 & 2 & 1274.243 & 1304.137 & 1316 & 6 \\
16 & 0\textsuperscript{d} & 1326.675 & 1357.959 & 1424 & 17 \\
 & 1\textsuperscript{d} & 1364.150 & 1390.959 & 1217 & 6 \\
17 & 0 & 1397.233 & 1414.581 & 812 & 2 \\
\hline
Joint\textsuperscript{e} & -- & -46.441 & 1414.581 & 45762 & --
\enddata
\tablenotetext{\textrm{a}}{\bjdtdb--2455000.}
\tablenotetext{\textrm{b}}{Number of data points in the light curve after outlier removal and trimming.}
\tablenotetext{\textrm{c}}{Optimal order of the detrending polynomial used to model the systematics in the individual segment fits.}
\tablenotetext{\textrm{d}}{These segments display significant uncorrected systematics trends (particularly 15-0) and are not included in the joint fit.}
\tablenotetext{\textrm{e}}{The joint fit is carried out on the concatenated light curve constructed from the individual  systematics-removed segments.}
\end{deluxetable}

\subsection{Radial velocity measurements}\label{subsec:rv}

A total of eleven RV observations of KOI-964 were obtained over nine epochs using the High Resolution Echelle Spectrometer (HIRES) instrument at the Keck Observatory between UT 2014 Aug 2 and 2014 Sep 12. The first epoch consists of three consecutive exposures, and the RVs derived from those observations were combined into a single measurement. The HIRES spectra lie across three chips and span the wavelength regimes 364.3$-$479.5~nm, 497.7$-$642.1~nm, and 654.3$-$799~nm, respectively, resulting in 23 total echelle orders, each with a length of 4020~pixels. The first of these chips spans a wavelength range that contains many hydrogen features. The second of these chips contains the spectral features produced by the iodine wavelength calibration cell used to compute the wavelength solution.

The primary in the KOI-964 system is an A-type star, with a mass of $\sim$2.3~$M_{\odot}$ \citep{carter2011}. Extracting RV measurements for such a target presents two major challenges: first, the spectral features are wide and in most cases take up a large fraction of a single HIRES echelle order, resulting in complications for continuum normalization across those orders; second, since HIRES is not an environmentally stabilized spectrograph, the wavelength solution varies even over the course of a night. 

We derive relative RV values from the HIRES echelle spectra using the method of \citet{becker2015}. In summary, we utilize a simultaneous fit to all 23 HIRES orders using a two-dimensional model of the continuum level, which allows featureless orders to act as anchors in the fit. Eight of the orders have discernible spectral features, but the depth and width of these features prevent the determination of the true continuum level. We use the remaining featureless orders to constrain the overall continuum model. When a wavelength solution is not available for a given observation, we extrapolate from an existing solution taken closest in time to the target observation. These solutions can be generated when the iodine cell is used during an observation. The specifics of this extrapolation technique require extrapolating both between chips and between observations and are described in detail by \citet{becker2015}. We use a Markov Chain Monte Carlo routine to optimize the fit between our template spectrum (i.e., a smoothed version of the first acquired spectrum) and each of the subsequent spectra, simultaneously fitting the blaze function and the relative redshift. Unlike in \citet{becker2015}, we do not fit for the stellar $v \sin{i}$ due to the small number of data epochs and highly broadened spectra, which provide insufficient information content to fit the extra variable.  The fit to the full data cube produces relative RVs, where the RV of each observation is measured relative to the smoothed template spectrum. We finally apply a barycentric correction to the extracted RVs. The small number of features on the broadened spectra lead to relatively poor RV precision. The nine epochs of RVs obtained using this method are presented in Table~\ref{tab:rvs}. 

\section{Data Analysis}\label{sec:data}

\begin{deluxetable}{cr}
\tablewidth{0pc}
\tabletypesize{\scriptsize}
\tablecaption{
    Keck/HIRES Radial Velocity Measurements
    \label{tab:rvs}
}
\tablehead{
    \colhead{Epoch (\bjdtdb)} &
    \colhead{RV (km/s)} 
}
\startdata
2456872.02588 & $0.41\pm0.13$ \\
2456895.98257 & $14.56\pm1.25$ \\
2456906.86759 & $-19.57\pm2.93$ \\
2456907.88774 & $-10.24\pm4.24$ \\
2456908.86786 & $17.34\pm1.15$ \\
2456909.82046 & $-12.44\pm 1.94$ \\
2456910.92318 & $-6.38 \pm1.44$ \\
2456911.84000 & $16.73 \pm 1.34$ \\
2456912.94638 & $-2.86  \pm1.09$
\enddata
\end{deluxetable}


We analyze the KOI-964 phase curve using the ExoTEP pipeline --- a generalized Python-based tool in development for processing and analyzing the full range of time series datasets of relevance in exoplanet science \citep[e.g.,][]{benneke2019,wong2019}. ExoTEP provides a modular and customizable environment for handling datasets obtained from all space-based instruments currently or recently in operation, including \kepler, \textit{Hubble}, \textit{Spitzer}, and \textit{TESS}. The first application of ExoTEP to the study of full-orbit phase curves was published in \citet{shporer2019}.

\subsection{Eclipse model}\label{subsec:eclipse}

We model both transits (when the white dwarf passes in front of the larger primary) and secondary eclipses (when the white dwarf is occulted) using the \texttt{BATMAN} package \citep{kreidberg2015}. Throughout this work, we use the subscripts $a$ and $b$ to refer to the host star and the orbiting white dwarf, respectively. Under the parameterization scheme adopted by ExoTEP, the shape and timing of these eclipse light curves $\lambda(t)$ are determined by the radius ratio $R_{b}/R_{a}$, the relative brightness of the secondary $f_{b}$, the impact parameter $b$, the scaled orbital semi-major axis $a/R_{a}$, a reference mid-transit time $T_{0}$, the orbital period $P$, the orbital eccentricity $e$, and the argument of periastron $\omega$. The mid-transit time is determined for the zeroth epoch, which in the case of the joint fit is designated to be the transit event closest to the mean of the combined time series.

When generating the model eclipse light curves $\lambda(t)$ at the 30-minute cadence of the \kepler time series, we supersample the transit and secondary eclipse light curves in the vicinity of each event at 30-second intervals and average the finely-sampled fluxes in order to accurately compute the integrated flux at each exposure. We also account for the non-negligible relative brightness of the white dwarf and dilute the primary transits accordingly by the factor $(1+f_{b})$.

In the joint fit presented in this work, we allow all of the transit and orbital parameters to vary freely. On the other hand, when initially fitting for the individual segment light curves to optimize the systematics model (Section~\ref{subsec:systematics}), we fix $b$ and $a/R_a$ to the best-fit values from \citet{carter2011} and $e$ and $\omega$ to zero, since the short time baseline of each photometric series does not provide any strong constraints on orbital geometry, and the orbit of the system is consistent with circular (see Section~\ref{sec:res}).

For the transits, we employ a standard quadratic limb-darkening law to describe the radial brightness profile of the primary. In the joint fit, both coefficients $u_1$ and $u_2$ are allowed to vary, while in the individual segment fits, we fix them to the values from \citet{carter2011}: $u_1=0.20$ and $u_2=0.2964$. For the white dwarf secondary, \citet{carter2011} fixed the quadratic limb-darkening coefficient to zero and found that the remaining linear coefficient was largely unconstrained and consistent with zero. In our analysis, we fix both coefficients to zero and do not report any significant improvement to the quality of the joint fit when allowing those coefficients to vary.

\subsection{Phase curve model}\label{subsec:phase}

Following \citet{carter2011}, we model the out-of-occultation brightness variation as a third-order harmonic series in phase \citep[e.g.,][]{carter2011}:
\begin{equation}\label{eq:prefull}
\psi(t) = 1+\sum\limits_{k=1}^{3}A_{k}\sin(k\phi(t))+\sum\limits_{k=1}^{3}B_{k}\cos(k\phi(t)),\end{equation}
where $\phi(t) = 2\pi(t-T_{0})/P$, and we have normalized the flux such that the combined average brightness of both binary components is unity.

\subsection{Instrumental systematics}\label{subsec:systematics}

To model the instrumental systematics in each segment, we use a generalized polynomial in time:
\begin{equation} 
S^{\lbrace i\rbrace}_{n}(t) = \sum\limits_{k=0}^{n}c^{\lbrace i\rbrace}_{k}(t-t_{0})^{k}.
\end{equation}
Here, $t_{0}$ is the first time stamp in the light curve from segment $i$, and $n$ is the order of the polynomial model. The full phase curve model is given by
\begin{equation}\label{full}
F(t) = S^{\lbrace i\rbrace}_{n}(t) \times \lambda(t) \times \psi(t).
\end{equation}

To determine the optimal polynomial order for each segment, we carry out individual fits of each segment to the model in Equation~\eqref{full} and choose the polynomial order that minimizes the Bayesian information criterion (BIC): $\mathrm{BIC} \equiv k\log N-2\log L$, where $k$ is the number of free parameters in the fit, $N$ is the number of data points, and $L$ is the maximum log-likelihood. The optimal polynomial orders determined from our individual segment fits are listed in Table~\ref{tab:obs}. In all cases, when selecting polynomials of similar order, the best-fit astrophysical parameters do not vary by more than $0.3\sigma$. A compilation plot of the systematics-corrected light curves from all segments is provided in the Appendix. 

The residuals from the light curve fit for the first segment of Q15 (labeled as 15-0) show severe uncorrected systematics in the form of a quasi-periodic modulation with a characteristic frequency about 10\% higher than the orbital frequency. Close inspection of the binned residuals from the individual segment fits reveal several more instances of similar uncorrected systematics. The characteristic frequencies of these residual signals are close to one another, but not identical; the occurrences of these systematics are largely confined to individual segments, with adjacent segments showing clean, featureless residuals. These observations strongly rule out an astrophysical source of these additional photometric modulations, and they are likely to be instrumental systematics. In addition to 15-0, we find significant uncorrected systematics in the segments 3-0, 3-1, 5-0, 7-0, 7-1, 11-0, 11-1, 12-1, 16-0, and 16-1. We trim these segments in the joint light curve fit presented in this work. However, we find that including them in the photometric series does not affect the measured astrophysical parameters.

\subsection{Joint phase curve fit}\label{subsec:fit}

When carrying out the joint phase curve fit of the full \kepler photometric time series, we do not combine the uncorrected light curves and simultaneously fit the systematics model for all segments, since that would include over 200 systematics parameters and incur forbiddingly large computational overheads. Instead, we first remove $4\sigma$ outliers from the best-fit model for each segment and divide the photometric series by the best-fit systematics model from the individual analysis. Then, we concatenate the detrended flux arrays to form the combined light curve for our joint analysis. To empirically validate this approach, we have experimented with carrying out joint analyses of select subsets of data (e.g., Q2 and Q3 only) and comparing the results from (a) fits where we compute the full systematics model for all component segments and (b) fits where we use pre-detrended light curves and no systematics modeling. In all the cases we have tested, the astrophysical parameter estimates agree to within $0.2\sigma$. The combined light curve for our joint fit contains 34,293 data points.

In addition to the astrophysical parameters, we fit for a uniform per-point uncertainty $\sigma$ to ensure that the resultant reduced chi-squared value is near unity and self-consistently derive realistic uncertainties on the astrophysical parameters, given the intrinsic scatter of the light curves. The total number of free parameters in our joint phase curve analysis is 17.

The combined light curve spans 4 years and contains (either partially or in full) 229 transits and 237 secondary eclipses. The ExoTEP pipeline utilizes the affine-invariant Markov Chain Monte Carlo (MCMC) ensemble sampler \texttt{emcee} \citep{emcee} to simultaneously compute the posterior distributions of all free parameters. The number of walkers is equal to four times the number of free parameters, and each chain has a length of 30,000 steps. After inspecting the plotted chains by eye, we only use the last 40\% of each chain is extracted to generate the posterior distributions. We also apply the Gelman-Rubin convergence test \citep{gelmanrubin} to ensure that the diagnostic value $\hat{R}$ is well below 1.1.


\section{Results}\label{sec:res}

\subsection{Joint phase curve analysis}\label{subsec:joint}

\begin{deluxetable}{lll}
\tablewidth{0pc}
\tabletypesize{\scriptsize}
\tablecaption{
    Results of Joint Phase Curve Analysis
    \label{tab:modelparams}
}
\tablehead{
    \colhead{Parameter} &
    \colhead{Value}                     &
    \colhead{Error}    
}
\startdata
\sidehead{\textit{Fitted Parameters}}
$R_b/R_a$   \dotfill  & 0.080983 & 0.000077 \\
$f_b$  \dotfill  & 0.0189365 & $_{-0.0000041}^{+0.0000044}$ \\
$T_0$ (\bjdtdb)      \dotfill  & 2455662.252189 & $_{-0.000022}^{+0.000024}$ \\
$P$ (days)  \dotfill  & 3.273698741 & $_{-0.000000053}^{+0.000000054}$\\
$b$         \dotfill  & 0.6740 & $_{-0.0015}^{+0.0017}$\\
$a/R_a$     \dotfill  & 7.052 & $_{-0.015}^{+0.014}$\\
$e$   \dotfill  & 0.000271 & $_{-0.000029}^{+0.000153}$ \\
$\omega$ ($^{\circ}$) \dotfill  & $-$19.3 & $_{-36.1}^{+36.2}$ \\
$u_{1}$ \dotfill & 0.176 & $_{-0.031}^{+0.029}$ \\
$u_{2}$ \dotfill & 0.312 & $_{-0.034}^{+0.035}$ \\ 
$A_1$ (ppm) \dotfill  & 99.21  & $_{-0.86}^{+0.95}$ \\
$A_2$ (ppm) \dotfill  & $-$39.75   & $_{-0.98}^{+0.88}$ \\
$A_3$ (ppm) \dotfill  & 0.34  & $_{-0.90}^{+0.87}$ \\
$B_1$ (ppm) \dotfill  & 268.4 & 1.1 \\
$B_2$ (ppm) \dotfill  & $-$568.2 & 1.0 \\
$B_3$ (ppm) \dotfill  & $-$15.0  & $_{-1.1}^{+1.0}$\\
$\sigma$ (ppm) \dotfill  & 117.74 & $_{-0.47}^{+0.44}$ \\
\sidehead{\textit{Derived Parameters}} 
Transit depth (ppm)\tablenotemark{a}  \dotfill  & 6558 & 13\\
$i$ ($^{\circ}$)    \dotfill  & 84.515 &$_{-0.026}^{+0.023}$ \\
$e\cos\omega$ \dotfill &0.000239 & $_{-0.000012}^{+0.000011}$\\
$e\sin\omega$ \dotfill & $-$0.00008 & $_{-0.00026}^{+0.00016}$
\enddata
\tablenotetext{\textrm{a}}{Calculated as $(R_b/R_a)^2$.}
\end{deluxetable}

\begin{figure*}
\includegraphics[width=\linewidth]{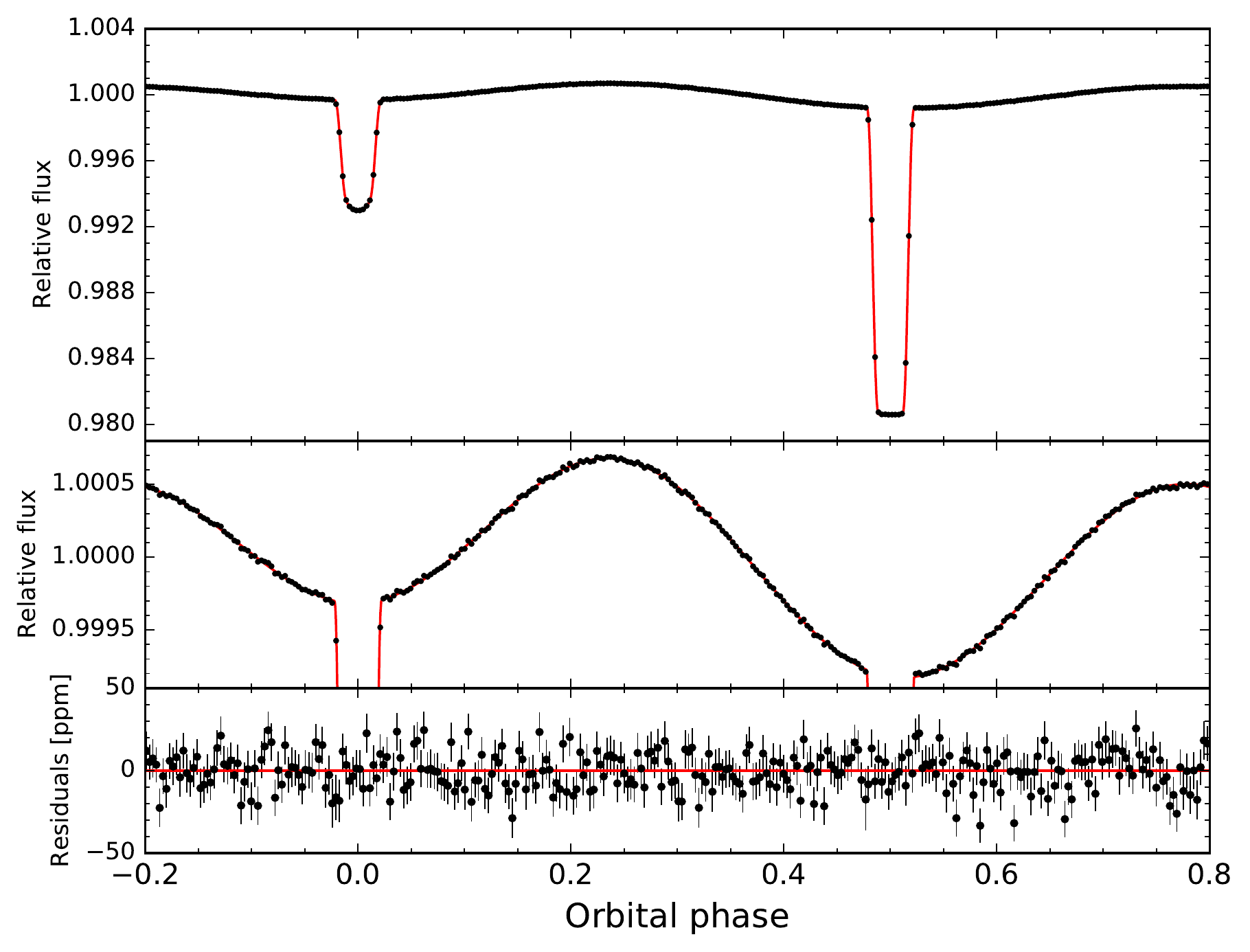}
\caption{Top panel: the phase-folded full light curve, after correcting for systematics trends, binned by 15-minute intervals (black points), along with the best-fit full phase curve model from our joint analysis (red line). Middle panel: same as top panel, but with an expanded vertical axis to detail the fitted phase curve modulation. Bottom panel: plot of the corresponding residuals from the best-fit model, in parts per million (ppm).}
\label{fig:bestfit}
\end{figure*}

The results of our joint analysis of all 18 \kepler quarters worth of data are listed in Table~\ref{tab:modelparams}.  We report the median and $1\sigma$ uncertainties for the astrophysical and noise parameters, as derived from the marginalized one-dimensional posterior distributions. The combined phase-folded light curve is plotted in Figure~\ref{fig:bestfit}, along with the best-fit full phase curve model.

When comparing with the results of \citet{carter2011}, we find that our astrophysical parameter estimates are at least 3.5~times more precise. Our inclination $i$ and $a/R_a$ estimates agree with the previously-published values at the $0.4\sigma$ level. The \citet{carter2011} analysis fixed the quadratic coefficient for the primary's limb-darkening law to the values calculated by \citet{sing} for a star with the stellar parameters of the primary in the KOI-964 system --- $u_2=0.2964$ --- while fitting for the linear coefficient and obtaining $u_1=0.20\pm0.02$. Our fitted values for $u_1$ and $u_2$ agree with these estimates at the $0.7\sigma$ and $0.5\sigma$ levels, respectively. The per-point uncertainty of 118~ppm that we obtain from our joint fit is about 10\% smaller than the value reported by \citet{carter2011}, indicating that data in more recent quarters have somewhat less scatter than the Q0 and Q1 photometry.

The period estimate from our joint analysis has an exquisite precision of 5~ms and differs from the value in \citet{carter2011} by $1.8\sigma$. The radius ratio $R_b/R_a$ is consistent with the previously-published value at $0.82\sigma$, while our secondary eclipse depth estimate $f_b$ is smaller by $3.1\sigma$. We have experimented with fitting only the data analyzed by \citet{carter2011} --- Q0 and Q1 --- and obtain $P$ and $f_b$ values that are consistent with their results at the $0.1\sigma$ and $0.6\sigma$ levels, respectively. This demonstrates that the addition of 16 more quarters of data has shifted the global estimate of the primary-secondary flux ratio in particular to a significantly different value.

The estimates we obtain for the main parameters of interest --- the phase curve harmonic amplitudes --- are consistent with the \citet{carter2011} values at better than the $1.1\sigma$ level. We detect the amplitudes of five harmonic terms at significance levels above $15\sigma$; they are, in order of decreasing amplitude and statistical significance, $\cos2\phi$, $\cos\phi$, $\sin\phi$, $\sin2\phi$, and $\cos3\phi$. The remaining harmonic, $\sin3\phi$, has an amplitude that is consistent with zero at the $0.4\sigma$ level. The astrophysical interpretation of these phase curve components will be discussed in detail in the following section.

One major discrepancy between our analysis and the results in \citet{carter2011} is the orbital eccentricity. They report an $e\cos\omega$ value of $0.0029\pm0.0005$, which translates to a significant delay in the midpoint of secondary eclipse relative to the expectation for a circular orbit: $\Delta t\equiv 2Pe\cos\omega/\pi \sim$~8.7~minutes (note: the value for the time delay reported in their work is erroneously off by a factor of 2). In our joint phase curve analysis, we obtain $e\cos\omega = 0.000239_{-0.000012}^{+0.000011}$, more than an order of magnitude smaller than the estimate in \citet{carter2011}. In a separate joint analysis of just Q0 and Q1 data, we find a similar value to the value from our full global fit, leading us to speculate that the reported $e\cos\omega$ value in \citet{carter2011} may be missing a zero after the decimal point. Our smaller updated constraint translates to a predicted secondary eclipse time delay of $43.0\pm2.1$~s. Assuming the orbital semi-major axis we derive from our stellar isochrone analysis and the fitted $a/R_{a}$ value (Section~\ref{subsec:params}) --- $0.0620\pm0.0044$~AU --- the relative light travel time delay between superior and inferior conjunctions is $61.9\pm4.4$~s, which is close ($<4\sigma$) to the secondary eclipse time delay predicted by our joint phase curve analysis. Therefore, we conclude that the orbit of the white dwarf is consistent with circular.

\begin{figure}
\includegraphics[width=\linewidth]{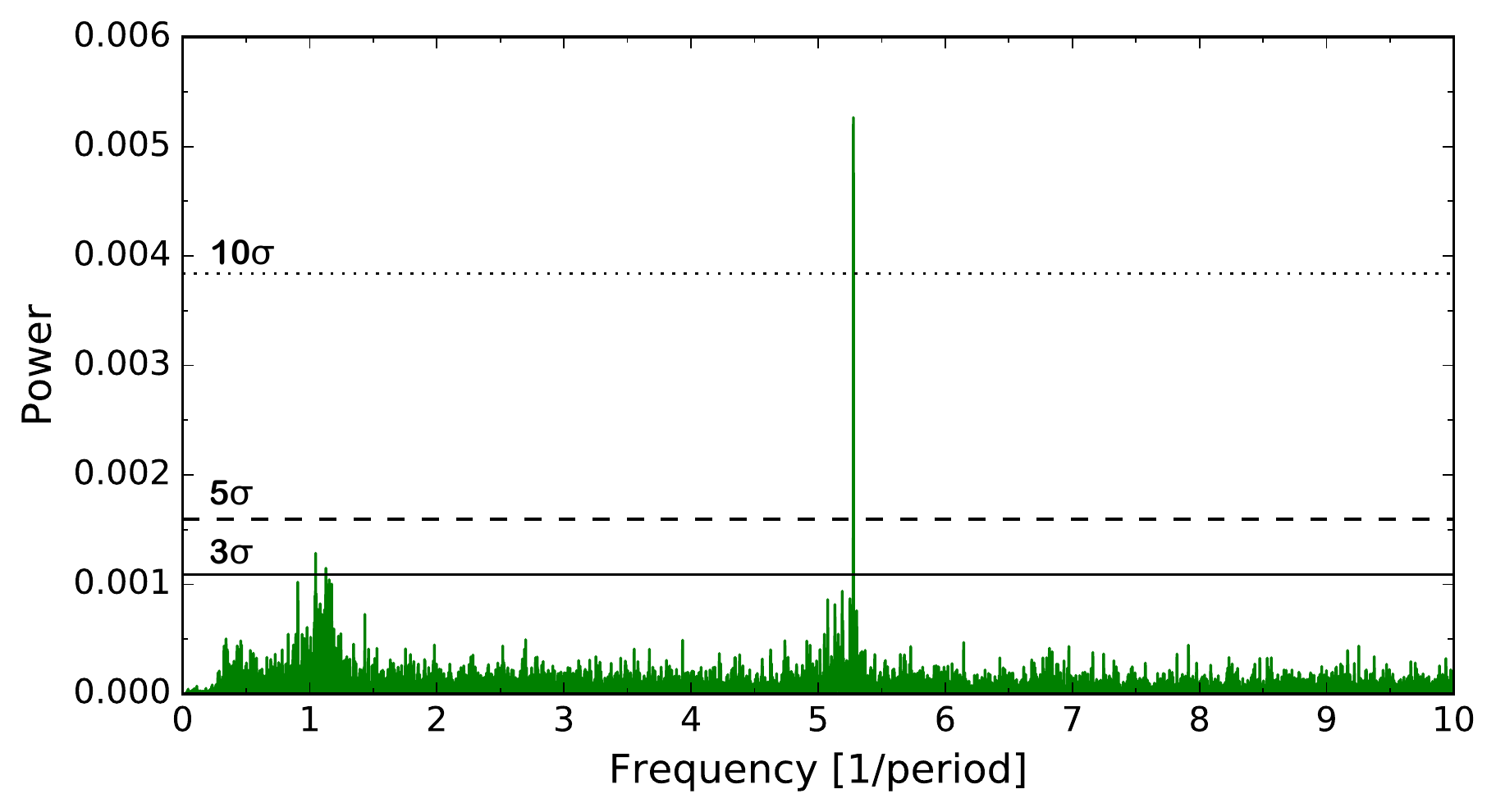}
\caption{Lomb-Scargle periodogram of the residuals from our joint phase curve fit. Horizontal lines indicate significance thresholds. There are no peaks at integer multiples of the orbital period. A strong peak corresponding to a period of around 0.62~days is evident and indicates stellar pulsations on the A-type host star in the KOI-964 system. Additional low-significance peaks near 1.1~period$^{-1}$ are attributable to residual systematics in the light curve and do not affect the astrophysical phase curve parameters measured in our fit.}
\label{fig:pgram}
\end{figure}

\subsection{Stellar pulsations}\label{subsec:pulsations}

Figure~\ref{fig:pgram} shows the Lomb-Scargle periodogram of the residuals from our best-fit phase curve model. No significant peaks at integer multiples of the orbital frequency are visible, demonstrating that our phase curve modeling, which includes Fourier terms up to third order, accounts for the full harmonic content in the astrophysical phase curve phased at the orbital period. 

There is a cluster of peaks in the power spectrum spanning frequencies somewhat higher than the orbital frequency. These signals are attributable to residual systematics that are not fully removed in the data trimming (see Section~\ref{sec:data}). When we include all segments into the joint fit, some peaks in this region of frequency space rise to 8--9$\sigma$ significance. With the trimming employed in the fit presented in this work, the noise contribution from these systematics is greatly reduced. Furthermore, the periodicities contributed by the systematics are distinct from the orbital frequency and all associated harmonics, and as such their presence does not affect the astrophysical phase curve parameters we measure in the joint fit.

A very strong single peak at a frequency of $\sim$5.28 period$^{-1}$ is evident in the periodogram, corresponding to a period of around 0.62~days. We have inspected all available \kepler light curves for stars within $\sim$1~arcmin of KOI-964 and do not find any periodicities at this frequency. Furthermore, the estimated relative contamination within the optimal extraction aperture for KOI-964 provided by the official \kepler data processing pipeline (given by the CROWDSAP keyword in the header) is 0\%. We conclude that this signal is most likely from the KOI-964 system. 

To further characterize this signal, we fit a simple sinusoidal function to the unbinned, non-phase-folded residual series:
\begin{equation}\label{pulse}
\Phi(t)=\beta\cos\left\lbrack 2\pi\frac{t-T_{0}-\tau}{\Pi}\right\rbrack.\end{equation}
Here $\beta$ and $\Pi$ are the semi-amplitude and period of the periodic signal, the zeropoint of the time series is set to the median mid-transit time from the joint fit $T_{0}=2455662.252189$, and $\tau$ represents the relative phase shift of this modulation. Our MCMC analysis yields $\Pi=0.620276\pm0.000011$~d, $\beta=12.1\pm0.9$~ppm, and $\tau=0.0453^{+0.0079}_{-0.0074}$~d.

The frequency of this observed periodic signal lies in the range spanned by $\gamma$ Dor pulsators \citep[e.g.,][]{guzik2000,balona2011}. Hundreds of $\gamma$ Dor pulsators have been discovered in the \kepler era, with typical pulsation frequencies smaller than 4~d$^{-1}$ and peaked around 1~d$^{-1}$ \citep[e.g.,][]{tkachenko2013,bradley2015}; some of the measured pulsation amplitudes from the \kepler sample are on the order of 10~ppm, consistent with the measured amplitude for the KOI-964 signal. However, $\gamma$ Dor pulsators tend to be late A-/F-type stars, with effective temperatures in the range 6000--8000~K \citep[e.g.,][]{tkachenko2013,bradley2015}, significantly cooler than KOI-964 ($T_{\mathrm{eff}}\sim 10000$~K). Therefore, the origin of the observed pulsation signal on KOI-964 remains uncertain.

\subsection{Results from RV analysis}\label{subsec:rvresults}
We use the \texttt{RadVel} package \citep{radvel} to model the orbit of the hot white dwarf based on the 9 Keck/HIRES RV measurements listed in Table~\ref{tab:rvs}. We place Gaussian priors on the orbital period ($P'$) and time of inferior conjunction (i.e., mid-transit time; $T'_{0}$) based on the median values and uncertainties from our joint phase curve fit (Table~\ref{tab:modelparams}). Here, the prime symbols ($'$) are used to distinguish parameters included in the RV analysis from the analogous parameters calculated in our joint phase curve fitting. We fit for the RV semi-amplitude $K_{\mathrm{RV}}$, along with the mean RV offset ($\gamma$) and jitter. The linear and second-order acceleration terms ($\dot{\gamma}$,$\ddot{\gamma}$) are fixed to zero. 

We run a fit that allows for independent constraints on orbital eccentricity ($e'$) and argument of periastron ($\omega'$), as well as a fit with eccentricity fixed to zero. The free-eccentricity RV fit yields $e'=0.082^{+0.079}_{-0.052}$, which is consistent with zero at the $1.6\sigma$ level and in line with the non-detection of significant secondary eclipse timing offset in our joint phase curve fit. When fitting an RV model with orbital eccentricity fixed to zero, we obtain a similar BIC value to the free-eccentricity fit, while the Akaike information criterion with small sample size correction (AICc) strongly favors the zero-eccentricity model ($\Delta\mathrm{AICc}=90.83$). Therefore, we present the RV fit results assuming a circular orbit in this work.

\begin{deluxetable}{lll}
\tablewidth{0pc}
\tabletypesize{\scriptsize}
\tablecaption{
    Results of RV Analysis
    \label{tab:rvfit}
}
\tablehead{
    \colhead{Parameter} &
    \colhead{Value}  &
    \colhead{Error}    
}
\startdata
$P'$ (days)\tablenotemark{a} \dotfill & 3.273698811 & $_{-0.000000048}^{+0.000000049}$ \\
$T'_{0}$ (\bjdtdb)\tablenotemark{a}      \dotfill  & 2455665.525884 & 0.000022 \\
$e'$ \dotfill & $\equiv$ 0.0 & --- \\
$K_{\mathrm{RV}}$ (km/s) \dotfill & 18.5 & $^{+2.0}_{-1.8}$ \\
$\gamma$ (km/s)\tablenotemark{b} \dotfill & $-$1.3 & 1.3 \\
jitter (km/s) \dotfill  & 3.2 & $_{-1.0}^{+1.5}$ 
\enddata
\tablenotetext{\textrm{a}}{Constrained by Gaussian priors derived from joint phase curve fit estimates (Table~\ref{tab:modelparams}).}
\tablenotetext{\textrm{b}}{Linear and second-order RV acceleration terms fixed to zero.}
\end{deluxetable}

\begin{figure}
\includegraphics[width=\linewidth]{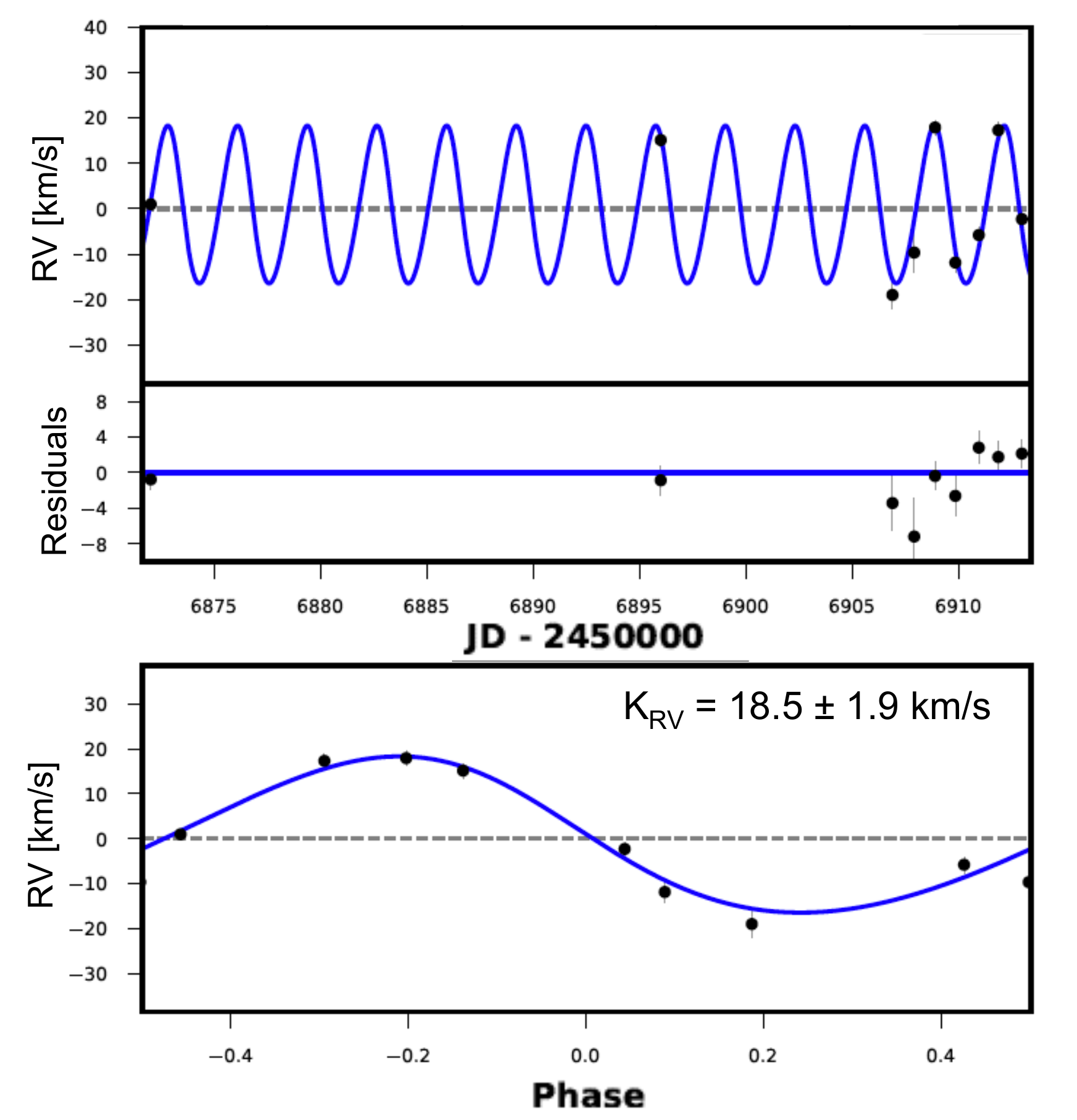}
\caption{Top panel: radial velocity (RV) measurements from Keck/HIRES (black points) and the best-fit RV curve (blue line). Middle panel: corresponding residuals from the fit. Bottom panel: phased RV curve, along with the derived RV semi-amplitude $K_{\mathrm{RV}}$).}
\label{fig:rv}
\end{figure}

The results of our RV fitting are shown in Table~\ref{tab:rvfit} and Figure~\ref{fig:rv}. The RV variation, with a fitted semi-amplitude of $K_{\mathrm{RV}}=18.5^{+2.0}_{-1.8}$~km/s, is detected at the $10.3\sigma$ level.

\subsection{Updated stellar parameter estimates}\label{subsec:params}

In order to calculate the mass, radius, and orbital semi-major axis of the transiting white dwarf from the measured RV variation and fitted orbital parameters, we must first obtain a reliable estimate of the host star's mass and radius. 

We follow a methodology similar to the one described in \citet{berger2018} and utilize the stellar classification package \texttt{isoclassify} \citep{isoclassify}. For input, we combine the parallax measurement from \textit{Gaia} Data Release 2 \citep[$0.4257\pm0.0348$~mas;][]{lindegren2018} with the 2MASS $K$-band magnitude ($13.056\pm0.029$~mag) and SDSS $g$-band magnitude ($13.000\pm0.020$~mag). We correct the \textit{Gaia} parallax with a positive offset of 0.03~mas \citep{berger2018,lindegren2018}.

The \texttt{isoclassify} routine combines the distance calculated from the input parallax with the extinction derived from a 3D dust reddening map and interpolated reddening vectors listed in \citet{green2018} to estimate the absolute magnitude of the host star. Using the ``grid method'', as opposed to the ``direct method'' utilized in \citet{berger2018}, then allows for the posteriors of the host star's properties to be self-consistently computed through comparisons with interpolated stellar isochrones from Modules for Experiments in Stellar Astrophysics Isochrones \& Stellar Tracks (MIST) grids \citep{choi2016}.

The presence of the binary companion can bias the retrieved parameters of the host star and lead to an overestimation in the host star's radius and temperature. In the \kepler bandpass, the white dwarf contributes roughly 2\% of the total luminosity in the system. As a first-order correction for the contamination, we first use \texttt{isoclassify} on the uncorrected magnitudes to derive the host star's temperature and radius. We then approximate the spectra of the host star and white dwarf companion as blackbodies and utilize the measured flux ratio $f_{b}$ in the \kepler bandpass and the measured radius ratio $R_{b}/R_{a}$ from the joint light curve fit to compute the white dwarf's effective temperature (Equation~\eqref{tempratio}). Lastly, while fixing the component radii and the computed white dwarf flux in the \kepler bandpass, we adjust the host star's temperature to match the flux ratio $f_b$. By calculating the difference in the host star's integrated flux in $g$- and $K$-band between the initial and adjusted blackbody spectra, we obtain magnitude corrections of $\Delta g = +0.026$~mag and $\Delta K= + 0.013$~mag.

Table~\ref{tab:properties} lists the estimates of the host star's properties we obtain from \texttt{isoclassify}, using the corrected magnitudes. All of these values are consistent to within $1\sigma$ with the corresponding values derived from using the uncorrected magnitudes. Our value for $R_{a}$ is in good agreement with the values computed by \citet{berger2018} using the ``direct method'', while the stellar mass $M_{a}$ is consistent with the value in \citet{mathur2017} --- $2.760 ^{+0.289}_{-0.674}$~$M_{\Sun}$ --- which was derived without including the \textit{Gaia} parallax.

Having estimated the host star's radius $R_{a}$, we use the best-fit values for $R_{b}/R_{a}$ and $a/R_{a}$ from the joint phase curve analysis (Table~\ref{tab:modelparams}) to calculate the radius and orbital semi-major axis of the white dwarf companion: $R_{b}=0.153\pm 0.011$~$R_{\Sun}$ and $a=0.0620\pm0.0044$~AU.

We separately calculate the mass ratio $q\equiv M_{b}/M_{a}$ and the white dwarf's mass $M_{b}$ using the measured RV semi-amplitude $K_{\mathrm{RV}}$ and the primary's mass derived above.
In the case of a circular orbit, the quantities $q$ and $M_{b}$ are related to $K_{\mathrm{RV}}$ by the following expression:
\begin{align}\label{rv}K_{\mathrm{RV}} &= q\sin i \left\lbrack\frac{2\pi G} {P}\frac{M_a}{(1+q)^2}\right\rbrack^{1/3} \notag \\ &= M_b\sin i \left\lbrack\frac{2\pi G}{P}\frac{1}{(M_a+M_b)^2}\right\rbrack^{1/3}.\end{align}

To construct the posteriors of $q$ and $M_{b}$, we randomly sample combinations of ($P$,$i$,$M_{a}$,$K_{\mathrm{RV}}$) from the respective posteriors and numerically solve for $q$ and $M_b$. The resultant estimates are $q = 0.106\pm0.012$ and $M_{b}=0.236^{+0.028}_{-0.027}$~$M_{\Sun}$. Our deduced properties of the white dwarf companion are summarized in Table~\ref{tab:properties}.

The density of the host star $\rho_a$ can be calculated in two ways. The first method is direct, $\rho_a=M_a/(4\pi R_a^3/3)$, while the second method utilizes Kepler's third law and the mass ratio $q$ to infer $\rho_a$ from the orbital period and semi-major axis:
\begin{equation}\label{density}\rho_a=\frac{3\pi}{GP^2}\frac{1}{1+q}\left(\frac{a}{R_a}\right)^3.\end{equation}
Using these two methods, we obtain the statistically consistent estimates $0.450^{+0.082}_{-0.079}$~g/cc and $0.5597^{+0.0071}_{-0.0069}$~g/cc. The second method relies on quantities with significantly smaller relative uncertainties and therefore yields the more precise density estimate, which we take and list in Table~\ref{tab:properties}. For the white dwarf's density, we can only utilize the direct method.

\begin{deluxetable}{lll}
\tablewidth{0pc}
\tabletypesize{\scriptsize}
\tablecaption{
    KOI-964 System Properties
    \label{tab:properties}
}
\tablehead{
    \colhead{Parameter} &
    \colhead{Value}                     &
    \colhead{Error}    
}
\startdata
\sidehead{\textit{(1) A-type host star}\tablenotemark{a}}
$R_{a}$ ($R_{\Sun}$) \dotfill & 1.89 & $^{+0.14}_{-0.13}$\\
$M_{a}$ ($M_{\Sun}$) \dotfill & 2.23  & 0.12\\
$\rho_{a}$ (g/cc)\tablenotemark{b} \dotfill & 0.5597 & $^{+0.0071}_{-0.0069}$\\
$T_{\mathrm{eff},a}$ (K) \dotfill & 9940 & $^{+260}_{-230}$ \\
$\log{g}$ \dotfill & 4.226 & $^{+0.046}_{-0.054}$ \\
$\mathrm{[Fe/H]}$ \dotfill & $-0.09$ & $^{+0.15}_{-0.14}$\\
Luminosity, $L_1$~($L_{\Sun}$) \dotfill & 31.7 & $^{+5.8}_{-4.5}$\\
Distance, $d$ (pc) \dotfill & 2220 & $^{+170}_{-140}$\\
Age (Gyr) \dotfill & 0.21 & $^{+0.11}_{-0.08}$ \\
\hline
\sidehead{\textit{(2) Transiting white dwarf companion\tablenotemark{b}}} 
$R_{b}$ ($R_{\mathrm{\Sun}}$)  \dotfill  & 0.153 & 0.011 \\
$q\equiv M_b/M_a$ \dotfill & 0.106 & 0.012 \\
$M_{b}$ ($M_{\mathrm{\Sun}}$)   \dotfill  & 0.236 & $^{+0.028}_{-0.027}$\\
$\rho_{b}$ (g/cc) \dotfill & 93 & 23\\
$T_{\mathrm{eff},b}$ (K) \dotfill & 15080 & 400\\
$a$ (AU) \dotfill & 0.0620 & 0.0044 
\enddata
\tablenotetext{\textrm{a}}{Computed with \texttt{isoclassify} using \textit{Gaia} parallax, 2MASS $K$-band photometry, and the SDSS \textit{g}-band magnitude.}
\tablenotetext{\textrm{b}}{Derived from $R_{a}$, $M_{a}$, and the results of our RV and joint phase curve analyses.}
\end{deluxetable}

Lastly, we estimate the temperature of the white dwarf companion, $T_{\mathrm{eff},b}$. The secondary eclipse depth gives the relative fluxes of the two binary components and is related to their emission spectra $F_{\nu,a}$ and $F_{\nu,b}$ by 
\begin{equation}\label{tempratio}f_b = \left(\frac{R_b}{R_a}\right)^2 \frac{\langle F_{\nu,b}\rangle}{\langle F_{\nu,a}\rangle},\end{equation}
where the spectra are averaged over the \kepler bandpass, weighted at each frequency by the corresponding value in the \kepler response function. We model the host star's emission using \texttt{PHOENIX} stellar spectra \citep{phoenix}, while the white dwarf's flux is represented as a simple blackbody. 

To facilitate with estimating $T_{\mathrm{eff},b}$ and reliably propagating the uncertainties in stellar properties to the uncertainty in $T_{\mathrm{eff},b}$, we derive empirical analytic expressions for $\langle F_{\nu,a}\rangle$ and $\langle F_{\nu,b}\rangle$ as functions of ($T_{\mathrm{eff},a}$,[Fe/H],$\log{g}$) and $T_{\mathrm{eff},b}$, respectively. Using all available \texttt{PHOENIX} model stellar spectra spanning the ranges $T_{\mathrm{eff},a}=[8000,12000]$~K, $[\mathrm{Fe/H}]=[-1.0,0.5]$, and $\log{g}=[3.5,5.0]$, we compute $\langle F_{\nu,a}\rangle$ at each point in the three-dimensional grid and fit a generalized linear polynomial in the dependent variables. Similarly, we fit a cubic polynomial in $T_{\mathrm{eff},b}$ to the array of $\langle F_{\nu,b}\rangle$ values calculated for blackbody spectra across the temperature range $T_{\mathrm{eff},b}=[10000,20000]$~K at 50~K intervals. We obtain the following relationships (units of $\langle F_{\nu}\rangle$ are $10^{15}~\mathrm{erg/cm^{2}}$):
\begin{align}\label{flux1} \langle F_{\nu,a}\rangle &=4.709+10.525\tau_a\notag\\& \qquad+0.284\mathrm{[Fe/H]}-0.0081(\log{g}-4.0),\\
\label{flux2} \langle F_{\nu,b}\rangle &=4.856+13.934\tau_b+6.261\tau_b^2-1.864\tau_b^3,\end{align}
where $\tau_i \equiv \left( \frac{T_{\mathrm{eff},i}}{10000~\mathrm{K}} -1\right)$. Following a similar Monte Carlo sampling method used previously to compute the white dwarf's mass, we randomly sample from the posteriors of ($f_{b}$,$R_b/R_a$,$T_{\mathrm{eff},a}$,[Fe/H],$\log{g}$) and numerically compute the corresponding values of $T_{\mathrm{eff},b}$ to obtain $T_{\mathrm{eff},b}=15080\pm400$~K.

The large radius and high temperature of the white dwarf companion indicate that the object must be young and still cooling. When compared to the radius of a degenerate He star of the same mass, our measured value is roughly 8~times larger. As mentioned in \citet{carter2011}, cooling models of He white dwarfs show that lower mass objects with relatively H-rich atmospheres can remain bloated and hot for longer than their more massive, H-poor counterparts \citep{hansen1998,nelson2004}. Our measured mass $M_{b}=0.236^{+0.028}_{-0.027}$~$M_{\Sun}$ is consistent with the smaller of the two discrepant white dwarf masses that \citet{carter2011} derived from their phase curve analysis. They indicated that a H-rich 0.21~$M_{\Sun}$ white dwarf can remain large and hot for $\sim$0.15~Gyr. The young host star age we deduce from the isochrone analysis ($0.21^{+0.11}_{-0.08}$~Gyr) is therefore consistent with the predicted evolutionary tract of the hot white dwarf companion.

\section{Discussion}\label{sec:dis}

From our joint fit of the full \kepler light curve, we obtain an extremely precise measurement of the phase variation in the KOI-964 system. The various harmonic terms are linked to processes stemming from the mutual illumination and gravitational interaction between the binary components. The predicted amplitudes of these variations can be calculated from theoretical models using the fundamental properties of the system, such as the mass ratio, orbital semi-major axis, and stellar parameters.

\citet{carter2011} carried out a retrieval analysis to estimate the stellar parameters and mass ratio using the results from their light curve fit and theoretical models for the phase curve terms. Having obtained the mass ratio and stellar properties independently from RV observations and isochrone fitting (Section~\ref{subsec:rvresults}), we are now in a position to calculate the predicted amplitudes of the phase curve terms using forward modeling and compare them with the observed values.

\subsection{Doppler boosting}\label{subsec:beaming}

As the two components of the binary system orbit around their center of mass, the apparent spectral intensity of both objects at a given wavelength modulates due to the periodic red- and blue-shifting of the spectra, as well as variations in the photon emission rate and light aberration \citep[e.g.,][]{shakura1987,loeb2003,zucker2007,shporer2010}. The composite effect is commonly referred to as Doppler boosting. This temporal modulation is driven by the projected RVs of the two components, $v_{r,a}$ and $v_{r,b}$, and as such, the harmonic contribution of Doppler boosting to the phase curve is carried by the sine function of orbital phase. The amplitude $A_1^{\mathrm{DB}}$ of the Doppler boosting is related to the system properties through the following expression \citep[e.g.,][]{loeb2003,shporer2010, carter2011}:
\begin{align}\label{db}A_{1}^{\mathrm{DB}} &= \alpha_{a}\left(\frac{v_{r,a}}{c}\right)+\alpha_{b}f_b\left(\frac{v_{r,b}}{c}\right)\notag \\
&= \left(\frac{2\pi a}{Pc}\right)\left\lbrack\alpha_a\left( 1+\frac{1}{q}\right)^{-1}-\alpha_b (1+q)^{-1}f_b\right\rbrack.\end{align}
Here, $q$ is the mass ratio as defined previously, $f_b$ is the secondary eclipse depth, and $c$ is the speed of light.

The prefactors $\alpha_i$, in short, reflect the relative change in the objects' integrated fluxes through the observed bandpass due to Doppler shifting and depend on the shape of the emission spectra:
\begin{equation}\label{alpha}\alpha_i = 3-\left\langle \frac{d \log F_{\nu,i}}{d \log \nu}\right\rangle.\end{equation}
$F_{\nu}$ is the object's emission spectrum as a function of frequency $\nu$, and the derivative is averaged over the \kepler bandpass. 

We empirically model the dependence of $\alpha_a$ and $\alpha_b$ on stellar temperature, metallicity, and surface gravity using the same method described in Section~\ref{subsec:params}. In the case of $\alpha_a$, we find that including quadratic and cubic terms in temperature as well as a correlation term between temperature and surface gravity greatly improves the fit:
\begin{align}\label{a1}\alpha_a&= 2.077 - 1.543\tau_a +4.478\tau_a^2 -8.790\tau_a^3- 0.021\mathrm{[Fe/H]}\notag \\ &\qquad + 0.043(\log{g}-4.0)-0.311\tau_a(\log{g}-4.0),\\
\label{a2}\alpha_b&=2.635-1.852\tau_b+1.433\tau_b^2-0.502\tau_b^3,\end{align}
where, as before, $\tau_i \equiv \left( \frac{T_{\mathrm{eff},i}}{10000~\mathrm{K}} -1\right)$.

We calculate the predicted harmonic amplitude due to Doppler boosting using Equations~\eqref{db},\eqref{a1}, and \eqref{a2} by sampling the posteriors for the dependent variables --- $P$, $f_{b}$, $a$, $q$, $T_{\mathrm{eff},a}$, [Fe/H], $\log{g}$, and $T_{\mathrm{eff},b}$ (Tables~\ref{tab:modelparams} and \ref{tab:properties}). The resultant estimate is $A_{1}^{\mathrm{DB}}=114^{+18}_{-16}$~ppm, which is consistent with our best-fit phase curve amplitude ($A_1=99.21^{+0.95}_{-0.86}$~ppm) at the $0.9\sigma$ level. 

Among the three main processes that contribute to the out-of-eclipse phase curve modulation, Doppler boosting is solely responsible for the fundamental harmonic of the sine term. The consistency between the predicted value of $A_1$ computed using the RV-derived mass ratio $q$ and the observed photometric amplitude indicates that the formalism described above adequately captures the physical mechanisms that drive the Doppler boosting signal. We can then use the very precise measurement of $A_1$ from our joint phase curve fit and Equation~\eqref{a1} to derive an independent estimate of the mass ratio, $q^*$. Following a similar sampling method as before, we obtain $q^*=0.0932^{+0.0068}_{-0.0058}$, which is consistent with the value derived from our RV analysis (Table~\ref{tab:properties}) at the $0.9\sigma$ level and twice as precise.

\subsection{Mutual illumination}\label{subsec:illumination}

In a binary system of two self-luminous objects, the radiation emitted by one component is incident on the other and is subsequently scattered or absorbed and reemitted. The regions near the sub-stellar points on the mutually facing hemispheres are therefore expected to be brighter than the other regions. Over the course of an orbit, the viewing phase of the secondary and the position of the illuminated region on the primary both change, imparting a periodic brightness variation to the phase curve. The maximum illumination of the primary is observed during mid-transit, while the primary-facing hemisphere of the secondary is fully oriented toward the observer during mid-eclipse. Both of these variations contribute primarily to the fundamental cosine mode in the phase curve harmonic series, albeit with opposite signs.

Under the assumption of radiative equilibrium, i.e., when all incident radiation is reemitted at the effective temperature of the illuminated object, the amplitudes of the mutual illumination modulation are given by \citep{kopal1959}
\begin{align}
\label{illum1}B_{1}^{\mathrm{ILL}} =& \frac{17}{16}\left(\frac{R_a}{a}\right)^{2}\left\lbrack\frac{1}{4}\frac{R_a}{a}+\frac{1}{3}\right\rbrack\left(\frac{T_{\mathrm{eff},b}}{T_{\mathrm{eff},a}}\right)^4\left(\frac{R_b}{R_a}\right)^2\notag\\
&-\frac{17}{16}\left(\frac{R_b}{a}\right)^{2}\left\lbrack\frac{1}{4}\frac{R_b}{a}+\frac{1}{3}\right\rbrack\frac{\beta_b}{\beta_a},\\
\label{illum2}B_{2}^{\mathrm{ILL}} =& \frac{17}{16}\left(\frac{R_a}{a}\right)^{2}\left\lbrack\frac{3}{16}\frac{R_a}{a}+\frac{16}{27\pi^2}\right\rbrack\left(\frac{T_{\mathrm{eff},b}}{T_{\mathrm{eff},a}}\right)^4\left(\frac{R_b}{R_a}\right)^2\notag\\
&-\frac{17}{16}\left(\frac{R_b}{a}\right)^{2}\left\lbrack\frac{3}{16}\frac{R_b}{a}+\frac{16}{27\pi^2}\right\rbrack\frac{\beta_b}{\beta_a},
\end{align}
where the bolometric correction to optical wavelengths $\beta_i$ is approximated by \cite{allen1964}
\begin{equation}\label{bc}\log{\beta_i} \simeq 17.0-4\log{T_{\mathrm{eff},i}}-\frac{11600~\mathrm{K}}{T_{\mathrm{eff},i}}.\end{equation}

With the measured and derived values for $a/R_a$, $R_b/R_a$, $T_{\mathrm{eff},a}$, and $T_{\mathrm{eff},b}$ as input, Equations~\eqref{illum1}--\eqref{bc} predict the following amplitudes for the mutual illumination variation: $B_{1}^{\mathrm{ILL}}=261^{+44}_{-39}$~ppm and $B_{2}^{\mathrm{ILL}}=62.0^{+10.3}_{-9.0}$~ppm.


\begin{table*}[t!]
\begin{center}
\caption{Predicted and Observed Phase Curve Amplitudes (in ppm)} \label{tab:amp} 

\begin{tabular}{c|cccc|ccc}
\hline
\hline
& Doppler  & Mutual & Ellipsoidal & Ellipsoidal & Predicted & Predicted &\textbf{Observed} \\
& boosting & illumination & distortion & distortion (with $q^*$)\tablenotemark{a}& & (with $q^*$)\tablenotemark{a} & \\
\hline
$A_1$ & $114^{+18}_{-16}$ & --- & --- & ---& $114^{+18}_{-16}$& $\equiv99.21^{+0.95}_{-0.86}$ & $\mathbf{99.21^{+0.95}_{-0.86}}$\\
$A_2$ & --- & --- & --- & --- & 0 & 0 &$\mathbf{-39.75^{+0.88}_{-0.98}}$\\
$B_1$ & --- & $261^{+44}_{-39}$ & $-8.9\pm1.0$ & $-7.80^{+0.56}_{-0.58}$& $252\pm42$ & $253\pm42$ & $\mathbf{268.4\pm1.1}$\\
$B_2$ & --- & $62.0^{+10.3}_{-9.0}$ & $-451\pm51$& $-396^{+27}_{-28}$ & $-389\pm52$ & $-334\pm29$ & $\mathbf{-568.2\pm1.0}$\\
$B_3$ & --- & --- & $-15.3\pm1.8$& $-13.5\pm1.0$ & $-15.3\pm1.8$& $-13.5\pm1.0$ & $\mathbf{-15.0^{+1.0}_{-1.1}}$\\
\hline
\end{tabular}
\end{center}
\begin{tablenotes}
\footnotesize
\item \textsuperscript{a} These values are calculated using the mass ratio estimate $q^*$ inferred from the measured Doppler boosting phase curve amplitude $A_1=99.21^{+0.95}_{-0.86}$ (see Section~\ref{subsec:beaming}).
\end{tablenotes}

\end{table*}

\subsection{Ellipsoidal distortion}\label{subsec:ellipsoidal}

The gravitational interaction between the primary and the secondary produces prolate deviations from sphericity on both components, with the long axis of the resultant ellipsoidal shape lying very nearly along the line connecting the two components. The ellipsoidal distortion incurs variations in the sky-projected areas of both components as a function of orbital phase, yielding modulations in the apparent flux with maxima occurring during quadrature, i.e., a variation at the first harmonic of the cosine, $\cos{2\phi}$. 

The detailed physical formalism of the ellipsoidal modulation was derived in \citet{kopal1959} and can be described as a series of cosines, with the three leading-order terms having the following amplitudes:
\begin{align}\label{elp1}B_{1}^{\mathrm{ELP}} &= -Z^{(3)}q\left(\frac{R_a}{a}\right)^4(15\sin^3{i}-12\sin{i}),\\
\label{elp2}B_{2}^{\mathrm{ELP}} &= -Z^{(2)}q\left(\frac{R_a}{a}\right)^3\sin^2{i},\\
\label{elp3}B_{3}^{\mathrm{ELP}} &= -5Z^{(3)}q\left(\frac{R_a}{a}\right)^4\sin^3{i},\end{align}
where, in line with the notation of \citet{morris1985} and \citet{morris1993},
\begin{align}
\label{z2}Z^{(2)} &= \frac{3}{4(3-u_1)}\bigg\lbrack\frac{1}{5}(15+u_1)(1+\gamma_1)+\notag\\& \quad \frac{5}{16}(1-u_1)(3+\gamma_1)(6-7\sin^2{i})\left(\frac{R_a}{a}\right)^2\bigg\rbrack,\\
\label{z3}Z^{(3)} &= \frac{5u_1}{32(3-u_1)}(2+\gamma_1).
\end{align}
Here, $u_1$ and $\gamma_1$ are the limb-darkening and gravity-darkening coefficients, respectively, assuming a linear law. 

There is a straightforward relationship between $\gamma_1$ and $T_{\mathrm{eff},a}$ according to von Zeipel's law for early-type stars \citep{morris1985}:
\begin{equation}\label{gamma}\gamma_1=0.25\left\langle\frac{C/\lambda T_{\mathrm{eff},a}}{1-\exp(-C/\lambda T_{\mathrm{eff},a})}\right\rangle,\end{equation}
where $C=1.43879$~cm$\cdot$K, and $\lambda$ is wavelength. For the linear limb-darkening coefficient, we fit an analytic function to the values computed by \citet{sing} in the \kepler bandpass for the same grid of host star parameter values we used previously in Section~\ref{subsec:beaming}. Likewise, we calculate $\gamma_1$ for a range of host star temperatures and fit a polynomial through the resultant array of values. We obtain the following empirical relationships:
\begin{align}\label{g1}\gamma_1 &= 0.660-0.501\tau_a+0.590\tau_a^2-0.634\tau_a^3,\\
\label{u1}u_1 &= 0.445-0.325\tau_a+0.407\tau_a^2\notag\\
&\qquad+0.0014\mathrm{[Fe/H]}-0.0050(\log{g}-4.0).
\end{align}

Using Equations~\eqref{elp1}--\eqref{z3} and \eqref{g1}--\eqref{u1} and sampling the posteriors of ($T_{\mathrm{eff},a}$,[Fe/H],$\log{g}$,$q$,$a/R_a$,$i$) from Tables~\ref{tab:modelparams} and \ref{tab:properties}, we calculate the predicted ellipsoidal distortion amplitudes: $B_1^{\mathrm{ELP}}=-8.9\pm1.0$~ppm, $B_2^{\mathrm{ELP}}=-451\pm51$~ppm, and $B_3^{\mathrm{ELP}}=-15.3\pm1.8$~ppm.

We can instead use the more precise mass ratio $q^*=0.0932^{+0.0068}_{-0.0058}$ derived from the Doppler boosting term $A_1$ in the phase curve (see Section~\ref{subsec:beaming}) to produce a separate set of predicted ellipsoidal distortion amplitudes. An analogous calculation yields $B_1^{\mathrm{ELP},*}=-7.80^{+0.56}_{-0.58}$~ppm, $B_2^{\mathrm{ELP},*}=-396^{+27}_{-28}$~ppm, and $B_3^{\mathrm{ELP},*}=-13.5\pm1.0$~ppm.

\subsection{Inconsistency in first harmonic terms ($A_2$, $B_2$)}\label{subsec:inconsistency}

\begin{figure}
\includegraphics[width=\linewidth]{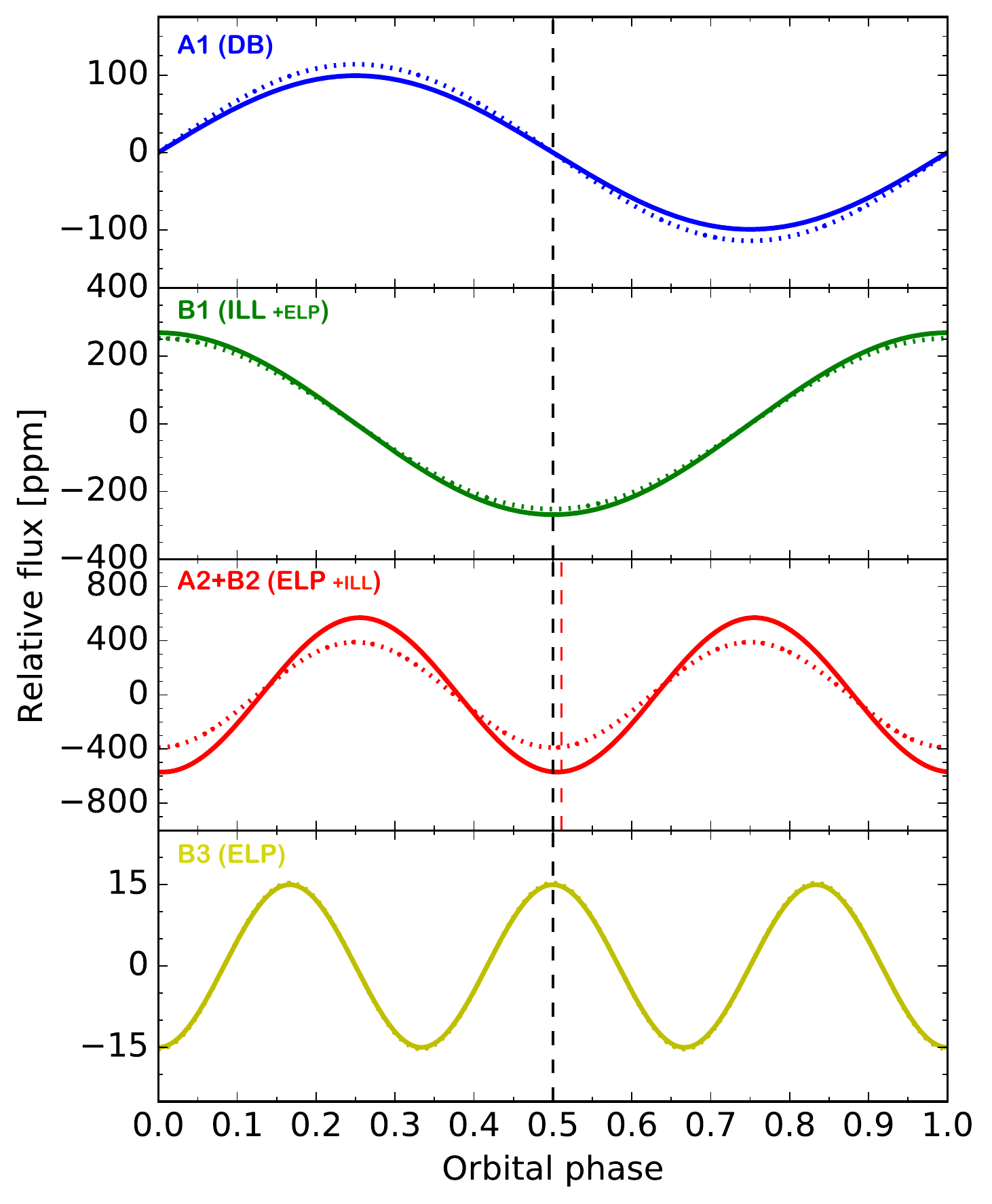}
\caption{Schematic of the photometric modulation components measured in our phase curve analysis of KOI-964, in parts per million (ppm) vs.~orbital phase. The solid curves show the measured signals at the corresponding term(s) of the Fourier series; the dotted curves show the predicted signals from theoretical modeling (see Table~\ref{tab:amp} and Sections~\ref{subsec:beaming}--\ref{subsec:ellipsoidal}). The physical process(es) that contribute to each component are indicated in parentheses, with the dominant contributor listed first. The black vertical dashed line denotes the mid-orbit phase (i.e., superior conjunction), while the red vertical dashed line illustrates the small phase lag ($\phi=4\overset{\circ}{.}00\pm0\overset{\circ}{.}09$) in the mid-orbit minimum of the combined first harmonic signal relative to expectation. Note the differences in vertical scale. The measured and predicted amplitudes are consistent to within $1\sigma$ in all cases, except for the combined first harmonic signal ($A_{2}+B_{2}$; third panel).}
\label{fig:comps}
\end{figure}

We summarize the measured values and theoretical predictions for the phase curve harmonic amplitudes in Table~\ref{tab:amp} and Figure~\ref{fig:comps}. We compute the total predicted values of all five harmonic terms with observed amplitudes significantly different from zero --- $A_1$, $A_2$, $B_1$, $B_2$, and $B_3$. For the ellipsoidal distortion amplitudes, the table lists values assuming both the RV-derived mass ratio $q$ and the mass ratio $q^*$ derived from the Doppler boosting term; in Figure~\ref{fig:comps} only the prediction calculated from the RV-derived mass ratio is shown. 

The observed amplitudes $A_1$, $B_1$, and $B_3$ all lie within $0.9\sigma$ of the predicted values. Doppler boosting is the only physical process that produces variation at the fundamental harmonic of the sine ($A_1$), while mutual illumination is the dominant contributor to the $B_1$ term. The consistency between the observed and predicted amplitudes indicates that the phase variations due to Doppler boosting and mutual illumination in the KOI-964 system are both well-described by the theoretical formalism presented in Sections~\ref{subsec:beaming} and \ref{subsec:illumination}.

The first harmonic of the cosine at the orbital phase ($B_2$ term) is a combination of contributions from mutual illumination and ellipsoidal distortion, with the latter being the predominant source of the variation. From Table~\ref{tab:amp}, we see that the total predicted amplitude, with the contribution from ellipsoidal distortion calculated using the mass ratio measured from RVs, differs from the observed value by $3.4\sigma$. This discrepancy was first noted in \citet{carter2011} based on the inconsistent mass ratios derived from Bayesian retrievals of the phase curve amplitudes including Doppler boosting or ellipsoidal distortion. We can perform an analogous consistency test by calculating the predicted $B_2$ value assuming the more precise mass ratio $q^*$ derived from the observed Doppler boosting variation. As shown in Table~\ref{tab:amp}, the inconsistency between this predicted value and the observed amplitude is more severe --- a $7.0\sigma$ discrepancy.


\subsubsection{Dynamical tide of the host star}\label{subsubsec:dynamical}

One hypothesis for the roughly $45\%$ underestimation of the ellipsoidal distortion amplitude from the modeling described in Section~\ref{subsec:ellipsoidal} relates to an oversimplification of the tidal dynamics on the host star. The formalism of \citet{kopal1959} only accounts for the equilibrium tide approximation; this approximation assumes that the distorted star maintains hydrostatic balance and thus ignores fluid inertia and the possibility of excited normal modes of oscillation, i.e., the dynamical tide. Detailed numerical modeling of the tidal response in stellar binaries has shown that the dynamical tide can contribute significantly to the observed flux perturbations, especially in the case of massive stars with largely radiative envelopes, such as KOI-964 \citep{pfahl2008}.

\citet{burkart2012} considered the surface flux perturbation due to equilibrium and dynamical tides on KOI-54 ($M=2.32\pm0.10\,M_{\Sun}$, $R = 2.19\pm0.03\,R_{\Sun}$), an A-type star similar to KOI-964 ($M=2.19\pm0.13\,M_{\Sun}$, $R = 1.92\pm0.13\,R_{\Sun}$). They showed that the flux perturbations caused by tidal distortion fall into three regimes based on the orbital period (see their Figure~6). For short periods ($P \lesssim 1\textrm{ day}$), the tide raised by the orbiting companion excites standing normal modes in the star, and the amplitude of the resultant flux perturbation is highly sensitive to resonances between the tidal forcing and the normal modes. For intermediate periods ($1 \lesssim P \lesssim 10~\textrm{days}$), the dynamical tide is strongly damped by rapid radiative diffusion near the surface, resulting in traveling waves rather than standing waves.  Although the resonances become severely attenuated in this regime, the amplitude of the flux perturbations are still significantly enhanced (by factors of order $\simeq 1-10$) relative to the prediction assuming only the equilibrium tide. Only for long periods ($P \gtrsim 10 \textrm{ days}$) does the equilibrium tide provide a good approximation of the flux perturbation. KOI-964 falls into the intermediate regime ($P = 3.2 \textrm{ days}$). It is therefore plausible that the influence of the dynamical tide is amplifying the flux variation stemming from the tidal response of the host star, thereby explaining the discrepancy between the observed amplitude and the theoretical value calculated in Section~\ref{subsec:ellipsoidal} for the equilibrium tide.

We detect a significant $41\sigma$ phase curve signal at the first harmonic of the sine ($A_2$). Such variation is not expected from any of the three processes modeled in Sections~\ref{subsec:beaming}--\ref{subsec:ellipsoidal}. An offset in the orientation of the primary star's ellipsoidal distortion relative to the orbiting white dwarf would manifest itself in our joint light curve fit as a non-zero $A_2$ amplitude. As shown in Table~\ref{tab:amp}, both $A_2$ and $B_2$ are negative, so the total first harmonic phase variation has extrema that occur later in the orbit than the corresponding extrema in the cosine-only curve. Combining the $A_2$ and $B_2$ terms into a single cosine signal yields a small but statistically significant phase lag of $\phi=4\overset{\circ}{.}00\pm0\overset{\circ}{.}09$. This phase lag is illustrated in Figure~\ref{fig:comps}.

Here, once again, the excitation of the dynamical tide and the relationship between the orbital frequency and the characteristic harmonic frequencies of the stellar oscillations may explain the observed behavior. \citet{burkart2012} showed that the relative phase between the maximum stellar surface displacement and the orbiting companion can vary from $-\frac{\pi}{2}$ to $+\frac{\pi}{2}$ depending on the proximity of the orbital period to resonances and the damping of the stellar oscillation modes: for near-resonance harmonics, the phase shift $|\phi|$ approaches $\frac{\pi}{2}$, while for cases in which the damping timescale is significantly longer than the orbital period, the phase shift is expected to be close to zero. As mentioned above, KOI-964 is expected to lie in a regime where resonances are strongly attenuated and traveling waves predominate at the stellar surface. Hence, the local phase of the stellar oscillations near the surface becomes important in describing the overall phase shift of the disk-averaged flux perturbations. Nevertheless, the formalism detailed in \citet{burkart2012} describers a plausible physical process by which the complex interactions between the host star's dynamical tide and the gravitational potential of orbiting white dwarf can induce a non-zero phase lag in the observed ellipsoidal distortion photometric modulation.

For a binary system where the companion's orbit is circular ($e=0$), the excitation of a dynamical tide on the host star requires non-synchronous rotation. Without a direct measurement of the stellar rotation frequency of the primary (see Section~\ref{subsec:rv}), we cannot determine whether the white dwarf's orbital period is synchronized to the host star's rotation period. Single A-type stars typically have rotational velocities exceeding 150~km/s, and the two other transiting hot white dwarf systems discovered by the \kepler Mission --- KOI-74 ($150\pm10$~km/s; \citealt{vankerkwijk2010,ehrenreich2011,bloemen2012}) and KOI-81 ($296\pm5$~km/s; \citealt{matson2015}) --- both have A- or B-type primary stars that rotate significantly faster than the orbital periods of their companions. Therefore, we posit that KOI-964 may also consist of a non-synchronous binary, for which interactions between the tidal bulge raised by the orbiting white dwarf and the host star's dynamical tide may manifest themselves in the measured photometric variability.

It is interesting to note here that KOI-74 also displays an ellipsoidal distortion modulation that deviates from the expected amplitude based on the equilibrium tide approximation \citep{rowe2010, vankerkwijk2010, ehrenreich2011, bloemen2012}. The KOI-74 system has an orbital period of 5.19~days and consists of a primary A-type star and a secondary white dwarf, similar to KOI-964, albeit with a smaller white dwarf. The discrepancy between the predicted mass ratio based on the Doppler boosting amplitude and that derived from the ellipsoidal distortion amplitude was noted already by \citet{vankerkwijk2010}. The mass ratio was directly measured using RV monitoring by \citet{ehrenreich2011} and \citet{bloemen2012}, who showed that it is consistent with the observed Doppler boosting photometric amplitude, while not consistent with the ellipsoidal distortion signal. However, in the case of KOI-74, the ellipsoidal distortion amplitude is \emph{smaller} than the predicted amplitude based on the equilibrium tide approximation, while for KOI-964 it is \emph{larger}. The differing behavior in these two systems shows that more work is needed to achieve a better understanding of the tidal distortion of hot stars.

More broadly, discrepancies between the mass ratios derived from measured Doppler boosting and ellipsoidal distortion amplitudes have been identified in a wide variety of systems, including both stellar binaries and star-planet systems, and with primary stars that range from cool Sun-like stars to hot giants, such as KOI-964 (for a detailed discussion, see Section~3.4 in \citealt{shporer2017}). For a few star-planet systems with cool star hosts, measurements of the mass ratios with RV monitoring have shown results that are consistent with the measured photometric ellipsoidal distortion amplitude but not the Doppler boosting signal. In these cases, the discrepancy has been attributed to a phase shift between the brightest region in the planet atmosphere and the substellar point \citep{shporer2015, parmentier2016}, which in turn biases the measured Doppler boosting amplitude and corresponding mass ratio prediction, since the mutual illumination and Doppler boosting phase components are both at the fundamental of the system's orbital period.

\subsubsection{Effects of rapid stellar rotation}\label{subsubsec:spinorbit}

Rapid rotation of the host star can affect the measured photometric variability. The rotational bulge induced by the star's spin produces large differences in surface temperature and surface gravity between the equator to the poles. As a result, the morphology of the tidal bulge is expected to deviate from the formalism of \citet{kopal1959}, which does not account for stellar rotation. \citet{vankerkwijk2010} modeled the effect of stellar rotation on the ellipsoidal distortion signal for KOI-74 and KOI-81 and found that varying the host star's rotation rate from synchronicity to 20~times faster than the orbital frequency increases the ellipsoidal distortion amplitude by up to a factor of $\sim$2. Therefore, the effect of the stellar rotational bulge may provide an explanation for the higher than expected $B_{2}$ value we measured from the phase curve analysis.

A non-zero spin-orbit misalignment introduces additional deviations in the ellipsoidal distortion modulation. While a binary companion with an equatorial orbit passes over regions of the star with the same surface gravity, this is not the case for a misaligned orbit. A spin-orbit misalignment would yield an additional photometric modulation signal at the first harmonic of the orbital period, since the average surface gravity across the tidal bulge comes to maximum and minimum twice during a single orbit. Crucially, because the three-dimensional orientation of KOI-964's spin axis is unconstrained, the relative phasing of this additional signal could vary from $-\pi$ to $+\pi$. Therefore, a misaligned orbit is able to produce both an amplitude deviation and a phase shift in the measured ellipsoidal distortion signal.

Likewise, spin-orbit misalignment can manifest itself in the measured mutual illumination signal. Because the surface temperature of a rapidly rotating star increases from the equator to the poles, the irradiation received by a misaligned orbiting companion varies across the orbit at the first harmonic of the orbital period. As in the case of the tidal bulge, the relative phase of this additional mutual illumination signal is unconstrained and can therefore produce a phase shift in the overall first harmonic photometric variation.

\section{Summary}\label{sec:sum}

We have presented here a phase curve analysis of KOI-964 incorporating all 18 quarters of \kepler data. The long observational baseline yields exquisite precision on the measured transit and secondary eclipse depths, as well as the phase curve amplitudes. The amplitudes of five sinusoidal phase curve harmonics were detected at higher than $15\sigma$ significance: $\cos2\phi$, $\cos\phi$, $\sin\phi$, $\sin2\phi$, and $\cos3\phi$. We also uncovered a stellar pulsation signal of indeterminate origin with a characteristic period of $0.620276\pm0.000011$~days and a peak-to-peak amplitude of $24\pm2$~ppm. Using stellar isochrone fitting and correcting for flux contamination by the white dwarf, we derived updated stellar parameters for both binary components and showed that the system is young ($0.21^{+0.11}_{-0.08}$~Gyr), consistent with the bloated size of the white dwarf. The results of our joint fit confirm the previous finding of \citet{carter2011} --- that the Doppler boosting and ellipsoidal distortion amplitudes predict inconsistent mass ratios. We obtained RV measurements of this system using the Keck/HIRES instrument and showed that the mass ratio calculated from the orbital RV signal is consistent with that predicted by the Doppler boosting amplitude. 

We hypothesize that the discrepancy between the observed and predicted ellipsoidal distortion modulation may stem from the non-convective nature of the hot A-type primary star, which allows the dynamical tide induced by the orbiting white dwarf companion to propagate to the stellar surface and interact with the equilibrium tide. The result is an amplification and phase shift of the ellipsoidal distortion photometric signal relative to the signal expected from assuming just the equilibrium tide. Another possible contributor to this discrepancy is rapid rotation of the host star, which incurs deviations in the surface gravity and temperature distributions across the stellar surface from the uniformity assumed in the standard tidal distortion model.

This study of KOI-964, along with previous studies of other binary systems with hot primary stars \citep[e.g.,][]{vankerkwijk2010, ehrenreich2011, bloemen2012}, shows that the tidal response of hot stars can deviate from the expected behavior under the assumption of equilibrium tides \citep{pfahl2008}. These findings serve as a cautionary tale against using the observed photometric ellipsoidal distortion amplitude to measure the mass ratio in systems with hot primary stars. The discrepancies between the expected and measured ellipsoidal distortion amplitudes in these systems also underscore the need for more careful and detailed modeling of the tidal response in hot stars.

\acknowledgments

This work includes data collected by the \kepler mission. Funding for the \kepler mission is provided by the NASA Science Mission directorate. We thank the California Planet Survey observers who helped to collect the Keck/HIRES data. Some of the data presented herein were obtained at the W. M. Keck Observatory, which is operated as a scientific partnership among the California Institute of Technology, the University of California and the National Aeronautics and Space Administration. The Observatory was made possible by the generous financial support of the W. M. Keck Foundation. The authors wish to recognize and acknowledge the very significant cultural role and reverence that the summit of Maunakea has always had within the indigenous Hawaiian community.  We are most fortunate to have the opportunity to conduct observations from this mountain. I.W.~and J.C.B.~are supported by Heising-Simons \textit{51 Pegasi b} postdoctoral fellowships.



\section*{Appendix}
The following plots show the results of individual segment light curve fitting. Each light curve is labeled by the \kepler quarter and segment number (see Table~\ref{tab:obs}). All the photometric series (black points) have been corrected by the corresponding best-fit systematics models. The best-fit phase curve models are overplotted in red. The bottom panels show the residuals from the best-fit models. Segments with significant residual systematics are marked with asterisks and are not included in our joint phase curve analysis. The first segment of quarter~15 (15-0) displays particularly large noise amplitudes. A machine readable table containing the data in these plots is available in the online version of this paper.

\begin{figure*}[h!]
\begin{center}
\includegraphics[width=\linewidth]{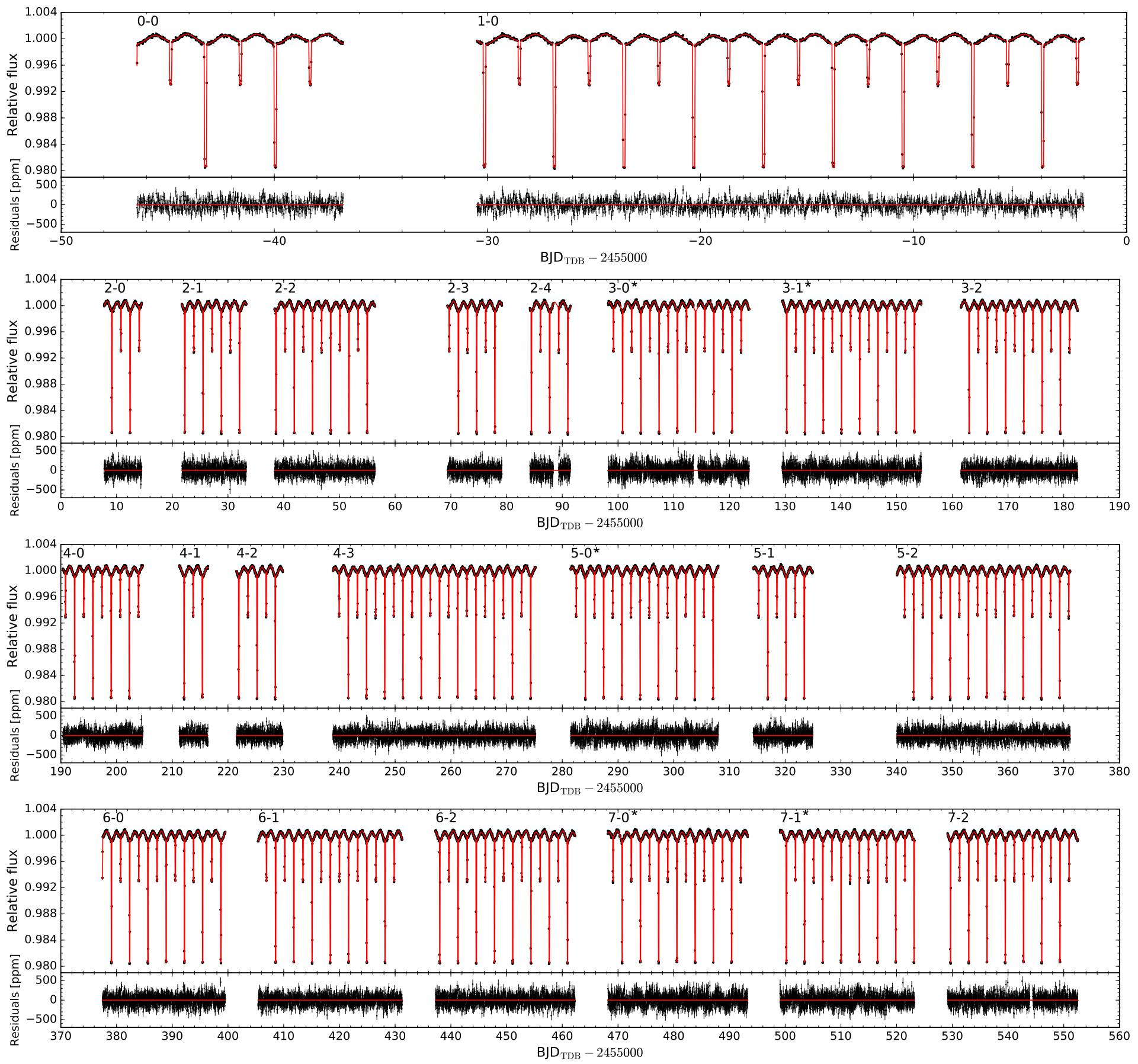}
\end{center}
\end{figure*}

\begin{figure*}[h!]
\begin{center}
\includegraphics[width=\linewidth]{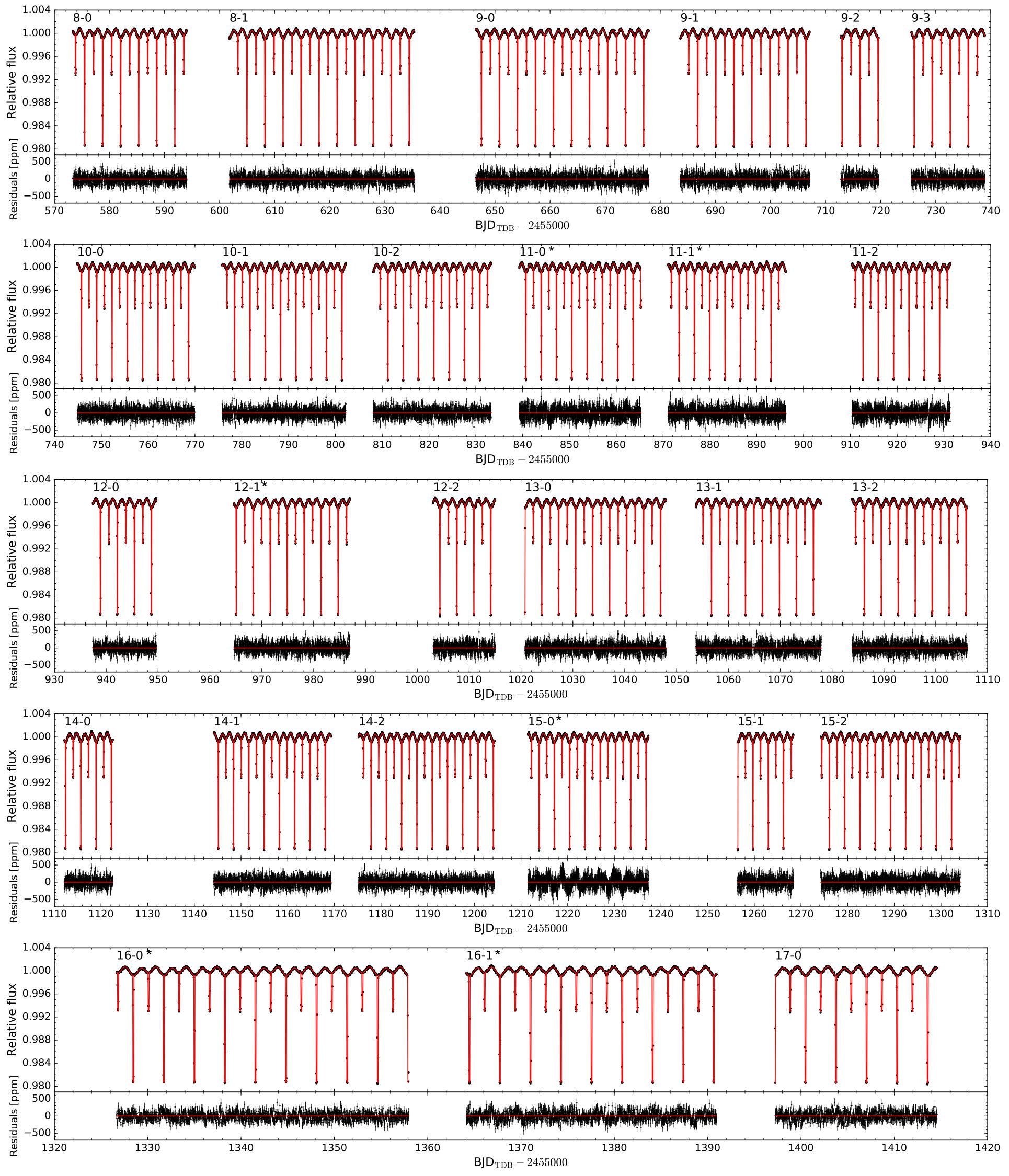}
\end{center}
\end{figure*}


\begin{thebibliography}{}
\tighten
\small


\bibitem[Allen(1964)]{allen1964}
Allen, C.~W. 1964, Astrophysical Quantities, The Athlone Press, University
of London



\bibitem[Balona et~al.(2011)]{balona2011}
Balona, L.~A., Guzik, J.~A., Uytterhoeven, K., et~al. 2011, \mnras, 415, 3531

\bibitem[Barclay et~al.(2012)]{barclay2012} Barclay, T., Huber, D., Rowe, J.~F., et~al. 2012, \apj, 761, 53

\bibitem[Becker et~al.(2015)]{becker2015} Becker, J.~C., Johnson, J.~A., Vanderburg, A., et~al. 2015, \apjs, 217, 29.

\bibitem[Berger et~al.(2018)]{berger2018}
Berger, T~A., Huber, D., Gaidos, E., \& van Saders, J.~L. 2018, \apj, 866, 99

\bibitem[Benneke et~al.(2019)]{benneke2019} Benneke, B., Knutson, H.~A., Lothringer, J. 2019, NatAs, 3, 813

\bibitem[Bloemen et~al.(2012)]{bloemen2012} Bloemen, S., Marsh, T.~R., Degroote, P., et~al. 2012, \mnras, 422, 2600 

\bibitem[Bradley et~sl.(2015)]{bradley2015}
Bradley, P.~A., Guzik, J.~A., Miles, L.~F., et~al. 2015, \aj, 149, 68

\bibitem[Burkart et~al.(2012)]{burkart2012} Burkart, J., Quataert, E., Arras, P., \& Weinberg, N.~N. 2012, \mnras, 421, 983


\bibitem[Carter et~al.(2011)]{carter2011}Carter, J.~A., Rappaport, S., \& Fabrycky, D. 2011, \apj, 728, 139

\bibitem[Choi et~al.(2016)]{choi2016}
Choi, J., Dotter, A., Conroy, C., et~al. 2016, \apj, 823, 102



\bibitem[Dai et~al.(2017)]{dai2017} Dai, F., Winn, J.~N., Yu, L., \& Albrecht, S. 2017, \aj, 153, 40 

\bibitem[Demory et~al.(2013)]{demory2013} Demory, B.-O., de Wit, J., Lewis, N., et~al.\ 2013, \apjl, 776, L25 


\bibitem[Ehrenreich et~al.(2011)]{ehrenreich2011} Ehrenreich, D., Lagrange, A.-M., Bouchy, F., et~al.\ 2011, \aap, 525, A85

\bibitem[Esteves et~al.(2013)]{esteves2013} Esteves, L.~J., De Mooij, E.~J.~W., \& Jayawardhana, R.\ 2013, \apj, 772, 51 

\bibitem[Esteves et~al.(2015)]{esteves2015} Esteves, L.~J., De Mooij, E.~J.~W., \& Jayawardhana, R.\ 2015, \apj, 804, 150 


\bibitem[Faigler \& Mazeh(2011)]{faigler2011} Faigler, S., \& Mazeh, T. 2011, \mnras, 415, 3921 

\bibitem[Faigler et~al.(2012)]{faigler2012} Faigler, S., Mazeh, T., Quinn, S.~N., Latham, D.~W., \& Tal-Or, L. 2012, \apj, 746, 185 

\bibitem[Faigler et~al.(2013)]{faigler2013} Faigler, S., Tal-Or, L., Mazeh, T., Latham, D.~W., \& Buchhave, L.~A. 2013, \apj, 771, 26 

\bibitem[Faigler \& Mazeh(2015)]{faigler2015} Faigler, S., \& Mazeh, T. 2015, \apj, 800, 73 

\bibitem[Foreman-Mackey et~al.(2013)]{emcee}
Foreman-Mackey, D., Hogg, D.~W., Lang, D., \& Goodman, J. 2013, \pasp, 125, 306

\bibitem[Fulton et~al.(2018)]{radvel}
Fulton, B.~J., Petigura, E.~A., Blunt, S., \& Sinukoff, E. 2018, \pasp, 130, 044504



\bibitem[Gelman \& Rubin(1992)]{gelmanrubin}
Gelman, A., \& Rubin, D.~B. 1992, StaSc, 7, 457

\bibitem[Green et~al.(2018)]{green2018}
Green, G.~M., Schlafly, E.~F., Finkbeiner, D., et~al 2018 \mnras, 478, 651

\bibitem[Guzik et~al.(2000)]{guzik2000}
Guzik, J.~A., Kaye, A.~B., Bradley, P.~A., Cox, A.~N., \& Neuforge, C. 2000, \apjl, 542, L57


\bibitem[Hansen \& Phinney(1998)]{hansen1998}
Hansen, B.~M.~S., \& Phinney, E.~S. 1998, \mnras, 294, 557

\bibitem[Huber et~al.(2014)]{huber2014}
Huber, D., Silva Aguirre, V., Matthews, J.~M., et~al. 2014, \apjs, 211, 2

\bibitem[Huber et~al.(2017)]{isoclassify}
Huber, D., Zinn, J., Bojsen-Hansen, M., et~al 2017, \apj, 844, 102

\bibitem[Husser et~al.(2013)]{phoenix}
Husser, T.-O., Wende-von Berg, S., Dreizler, S., et~al. 2013, \aap, 553, A6




\bibitem[Jackson et~al.(2012)]{jackson2012} Jackson, B.~K., Lewis, N.~K., Barnes, J.~W., et~al. 2012, \apj, 751, 112 

\bibitem[Jenkins \& Doyle(2003)]{jenkins2003} Jenkins, J.~M., \& Doyle, L.~R. 2003, \apj, 595, 429 

\bibitem[Jenkins et~al.(2010)]{jenkins2010} Jenkins, J.~M., Caldwell, D.~A., Chandrasekaran, H., et~al. 2010, \apjl, 713, L87 

\bibitem[Jenkins et~al.(2016)]{jenkins2016} Jenkins, J.~M., Twicken, J.~D., McCauliff, S., et~al. 2016, \procspie, 9913, 99133E

\bibitem[Jenkins et~al.(2017)]{jenkins2017} Jenkins, J.~M., Tenenbaum, P., \& Seader, S. 2017, Kepler Data Processing Handbook: Transiting Planet Search, Kepler Science Document, KSCI-19081-002


\bibitem[Kopal(1959)]{kopal1959}
Kopal, Z. 1959, Close Binary Systems, The International Astrophysics Series, Vol. 5 (London: Chapman \& Hall)

\bibitem[Kreidberg (2015)]{kreidberg2015}
Kreidberg, L. 2015, \pasp, 127, 1161




\bibitem[Loeb \& Gaudi(2003)]{loeb2003} Loeb, A., \& Gaudi, B.~S. 2003, \apjl, 588, L117 

\bibitem[Lindegren et~al.(2018)]{lindegren2018}
Lindegren, L., Hernandez, J., Bombrun, A., et~al 2018, \aap, 616, A2

\bibitem[Lomb(1976)]{lomb1976} Lomb, N.~R. 1976, \apss, 39, 447 


\bibitem[Mathur et~al.(2017)]{mathur2017}
Mathur, S., Huber, D., Batalha, N.~M., et~al 2017, \apjs, 229, 30

\bibitem[Matson et~al.(2015)]{matson2015}
Matson, R.~A., Gies, D.~R., Guo, Z., et~al. 2015, \apj, 806, 155

\bibitem[Morris(1985)]{morris1985} Morris, S.~L. 1985, \apj, 295, 143 

\bibitem[Morris \& Naftilan(1993)]{morris1993} Morris, S.~L., \& Naftilan, S.~A. 1993, \apj, 419, 344 


\bibitem[Nelson et~al.(2004)]{nelson2004}
Nelson, L.~A., Dubeau, E., \& MacCannell, K.~A. 2004, \apj, 616, 1124


\bibitem[Parmentier et al.(2016)]{parmentier2016} Parmentier, V., Fortney, J.~J., Showman, A.~P., Morley, C., \& Marley, M.~S.\ 2016, \apj, 828, 22 


\bibitem[Pfahl et~al.(2008)]{pfahl2008} Pfahl, E., Arras, P., \& Paxton, B. 2008, \apj, 679, 783 



\bibitem[Rappaport et~al.(2015)]{rappaport2015} Rappaport, S., Nelson, L., Levine, A., et~al. 2015, \apj, 803, 82 


\bibitem[Rowe et~al.(2010)]{rowe2010} Rowe, J.~F., Borucki, W.~J., Koch, D., et~al.\ 2010, \apjl, 713, L150


\bibitem[Scargle(1982)]{scargle1982} Scargle, J.~D. 1982, \apj, 263, 835 

\bibitem[Shakura \& Postnov(1987)]{shakura1987} Shakura, N.~I., \& Postnov, K.~A. 1987, \aap, 183, L21 


\bibitem[Shporer(2017)]{shporer2017} Shporer, A. 2017, \pasp, 129, 072001 

\bibitem[Shporer \& Hu(2015)]{shporer2015} Shporer, A., \& Hu, R. 2015, \aj, 150, 112 

\bibitem[Shporer et~al.(2011)]{shporer2011} Shporer, A., Jenkins, J.~M., Rowe, J.~F., et~al. 2011, \aj, 142, 195 

\bibitem[Shporer et~al.(2010)]{shporer2010} Shporer, A., Kaplan, D.~L., Steinfadt, J.~D.~R., et~al. 2010, \apjl, 725, L200 

\bibitem[Shporer et~al.(2014)]{shporer2014} Shporer, A., O'Rourke, J.~G., Knutson, H.~A., et~al. 2014, \apj, 788, 92 

\bibitem[Shporer et~al.(2019)]{shporer2019} Shporer, A., Wong, I., Huang, C.~X., et~al. 2019, \apj, 157, 178

\bibitem[Sing(2010)]{sing}
Sing, D.~K. 2010, \aap, 510, A21


\bibitem[Tal-Or et~al.(2015)]{talor2015} Tal-Or, L., Faigler, S., \& Mazeh, T. 2015, \aap, 580, A21 

\bibitem[Tkachenko et~al.(2013)]{tkachenko2013}
Tkachenko, A., Aerts, C., Yakushechkin, A., et~al. 2013, \aap, 556, A52



\bibitem[van Kerkwijk et~al.(2010)]{vankerkwijk2010} van Kerkwijk, M.~H., Rappaport, S.~A., Breton, R.~P., et~al. 2010, \apj, 715, 51 


\bibitem[Wong et~al.(2019)]{wong2019} Wong, I., Benneke, B., Gao, P., et~al. 2019, \apj, in revision



\bibitem[Zucker et~al.(2007)]{zucker2007} Zucker, S., Mazeh, T., \& Alexander, T. 2007, \apj, 670, 1326 


\end{thebibliography}
\end{document}